\documentclass[aip,jcp,twocolumn,reprint]{revtex4-1}

\usepackage{graphics,graphicx,amsfonts,amsmath,amsbsy,amssymb,color}
\usepackage{bm}
\usepackage[a4paper,vmargin={20mm,20mm},hmargin={20mm,10mm}]{geometry}
\usepackage{subfigure}
\usepackage{paralist}
\usepackage{bbold}

\def \Eh {{\textrm{E}_{\textrm{h}}}}

\newcommand{\bra}{\langle}
\newcommand{\ket}{\rangle}

\def \Nw {{N_{\textrm{w}}}}
\def \Eh {{E_{\textrm{h}}}}
\def \El {{\textrm{e}}}
\def \Or {{\textrm{o}}}

\begin{document}

\title{An efficient and accurate perturbative correction to initiator full configuration interaction quantum Monte Carlo}
\author{Nick~S.~Blunt}
\email{nicksblunt@gmail.com}
\affiliation{University Chemical Laboratory, Lensfield Road, Cambridge, CB2 1EW, United Kingdom}

\begin{abstract}
We present a perturbative correction within initiator full configuration interaction quantum Monte Carlo (i-FCIQMC). In the existing i-FCIQMC algorithm, a significant number of spawned walkers are discarded due to the initiator criteria. Here we show that these discarded walkers have a form that allows calculation of a second-order Epstein-Nesbet correction, that may be accumulated in a trivial and inexpensive manner, yet substantially improves i-FCIQMC results. The correction is applied to the Hubbard model, the uniform electron gas and molecular systems.
\end{abstract}

\date{\today}

\maketitle

Full configuration interaction quantum Monte Carlo (FCIQMC) was introduced by Booth, Thom and Alavi in 2009\cite{Booth2009}, and has since become a significant method in electronic structure theory for obtaining high-accuracy properties of challenging systems\cite{Booth2012, Shepherd2012_3, Thomas2015}. The method has led to the development of multiple other QMC-based approaches in the domain of quantum chemistry, including coupled cluster Monte Carlo (CCMC)\cite{Thom2012,Scott2017,Neufeld2017}, density matrix quantum Monte Carlo (DMQMC)\cite{Blunt2014, Malone2015, Malone2016} and model space quantum Monte Carlo (MSQMC)\cite{Ten-no2013, Ohtsuka2015, Ten-no2017}.

The efficiency of FCIQMC has been improved by several orders of magnitude since its introduction, primarily by a semi-stochastic adaptation\cite{Petruzielo2012, Blunt2015} and improved excitation generators\cite{Holmes2016_1}. Despite this, significant sources of inefficiency remain. One of the most notable such inefficiencies is due to the nature of spawning in the initiator adaptation to FCIQMC (i-FCIQMC)\cite{Cleland2010, Cleland2011}. In order to overcome the sign problem at low walker populations\cite{Spencer2012}, i-FCIQMC only allows spawned walkers to survive if they satisfy a set of criteria. Those that do not are removed from the simulation, along with significant information they contain about the space beyond the i-FCIQMC wave function.

Separately, selected configuration interaction (SCI) approaches are another important class of methods for obtaining FCI-level accuracy in challenging systems. SCI methods have existed for decades\cite{Huron1973, Buenker1974}, but have seen a particular renewal of interest in recent years\cite{Liu2016, Schriber2016, Tubman2016, Holmes2016_2}. SCI usually involves two stages. First a variational stage where a subspace of important determinants is generated, and in which the Hamiltonian eigenvalue problem is solved to give a zeroth-order energy and wave function. Second a perturbative correction is made, often of the second-order Epstein-Nesbet (EN2) type, which substantially corrects the zeroth-order energy. A semi-stochastic calculation of the EN2 correction was introduced into the heat-bath CI method (SHCI) by Sharma \emph{et al.}\cite{Sharma2017} A semi-stochastic EN2 calculation was also introduced by Garniron \emph{et al.},\cite{Garniron2017} which they applied to the CIPSI method\cite{Huron1973, Evangelisti1983, Scemama2013, Scemama2014, Caffarel2016}, although the approach presented is generally applicable. Extrapolation schemes have also been highly effective. With this, SCI methods have recently been used in several studies to obtain highly-accurate results for challenging systems, with modest computational resources\cite{Tubman2016, Garniron2017, Smith2017, Holmes2017, Chien2018, Scemama2018, Dash2018}.

Meanwhile, i-FCIQMC currently involves only the equivalent of the variational step, and yet with this alone has given accurate results for a large range of beyond-traditional-FCI systems. Given this accuracy, the question arises of whether a similar perturbative correction may be applied to i-FCIQMC, which could be extremely powerful. Here we show that such a correction is indeed possible, and that it can be built primarily from the information discarded in applying initiator criteria, and therefore already present in the simulation. As a result, this substantial improvement may be achieved in a natural and inexpensive manner.

We note that EN2 corrections have been applied to matrix product states in very recent work by Sharma\cite{Sharma2018} and also by Chan and co-workers\cite{Guo2018_1,Guo2018_2}, in separate studies. In the former, theory was also presented for applying an EN2 correction to more general non-linearly parameterized wave functions.

\emph{Theory:-} In FCIQMC, the ground-state wave function is converged upon by repeated application of a projection operator to some initial state, $\hat{P} = \mathbb{1} - \Delta\tau (\hat{H} - S\mathbb{1})$, for the Hamiltonian operator $\hat{H}$, a small parameter $\Delta \tau$, and a shift parameter $S$ for population control. This projection is performed stochastically, such that if the wave function at a given iteration is denoted $| \Psi (\tau) \ket$, the expected wave function at the subsequent iteration obeys $\textrm{E}[ \; | \Psi(\tau + \Delta \tau) \ket \; ] = \hat{P} | \Psi(\tau) \ket$, where $\textrm{E[\ldots]}$ denotes an expectation value\cite{fn2}, so that the correct projection is performed on average\cite{Booth2009, Spencer2012}. However, if $\hat{P}$ is applied without truncation then the FCIQMC algorithm quickly requires very large walker populations, making it impractical\cite{Spencer2012}.

Instead, FCIQMC studies have relied almost solely on the initiator adaptation (i-FCIQMC)\cite{Cleland2010, Cleland2011}. In i-FCIQMC the spawning of walkers is restricted, thus reducing the size of the space that can effectively be explored. The initiator rules are as follows. Determinants with more than $n_a$ walkers are defined as initiators, where $n_a$ is some small population threshold, typically $2$ or $3$. Initiators are allowed to spawn freely, with no truncation placed upon walkers spawned from them. In contrast, non-initiators may only spawn to already-occupied determinants, with an exception occurring if two or more spawning events occur to the same determinant in the same iteration, in which case the spawnings are allowed\cite{fn1}. Thus, i-FCIQMC effectively restricts application of $\hat{P}$ to within a subspace (albeit of a non-constant nature), and is therefore comparable to truncated-space methods (although it should be recognized that the effective space of i-FCIQMC is larger than the space of instantaneously occupied determinants).

Given this similarity with truncated-space methods, including the variational stage of SCI, we now consider how to apply a second-order perturbation correction in an analogous manner. Specifically, we use an Epstein-Nesbet (EN) partitioning, and therefore briefly describe EN perturbation theory, using the same notation as Sharma \emph{et al.}\cite{Sharma2017} In EN perturbation theory the space is split into a variational subspace, $\mathcal{V}$, spanned by determinants labelled by $| D_i \ket$ and $| D_j \ket$, and the rest of the space, spanned by determinants labelled $| D_a \ket$. The zeroth-order Hamiltonian is then defined as 
\begin{equation}
\hat{H}_0 = \sum_{ij \in \mathcal{V}} H_{ij} | D_i \ket \bra D_j| + \sum_{a \notin \mathcal{V}} H_{aa} | D_a \ket \bra D_a |,
\end{equation}
so that $\hat{H}_0$ contains the entire block of $\hat{H}$ within $\mathcal{V}$, while only consisting of the diagonal of $\hat{H}$ outside $\mathcal{V}$. As such, the ground state of $\sum_{ij\in\mathcal{V}} H_{ij} | D_i \ket \bra D_j|$, denoted $| \Psi_0 \ket = \sum_{i\in\mathcal{V}} c_i | D_i \ket$, is the zeroth-order wave function, and the corresponding eigenvalue is the zeroth-order energy, $E_0$. By standard perturbation theory the second-order energy correction may be calculated as
\begin{equation}
\Delta E_2 = \sum_{a\notin\mathcal{V}} \frac{(\sum_{i\in\mathcal{V}} H_{ai} c_i)^2}{E_0 - H_{aa}}.
\label{eq:E_2}
\end{equation}

We will now show that such a correction may be calculated in a simple manner when the zeroth-order wave function is sampled by FCIQMC. Before considering initiator FCIQMC, we first consider FCIQMC applied within a well-defined subspace (but \emph{without} initiator criteria). We again define this subspace as $\mathcal{V}$. One way to perform such a truncated FCIQMC calculation is by allowing generation of excitations to any determinant (connected by a single application of $\hat{H}$), and later removing any outside of $\mathcal{V}$. In the limit of large imaginary time, the FCIQMC wave function will sample the zeroth-order wave function, $| \Psi_0 \ket$ (with some non-unity normalization factor). Meanwhile the spawned vector sampled will be proportional to
\begin{equation}
\hat{P} | \Psi_0 \ket = | \Psi_0 \ket - \Delta \tau (\hat{H} - S\mathbb{1}) | \Psi_0 \ket,
\end{equation}
and so the expected contribution spawned onto determinants $|D_a\ket$ outside of $\mathcal{V}$ (which we label $S_a$), will obey $S_a \propto \bra D_a | \hat{P} | \Psi_0 \ket$, specifically
\begin{equation}
S_a \propto -\Delta\tau \sum_{i\in\mathcal{V}} H_{ai} c_i.
\label{eq:spawn_vec}
\end{equation}
It can therefore be seen that $(\sum_{i\in\mathcal{V}} H_{ai} c_i)^2$ may be estimated by $S_a^2/(\Delta \tau)^2$, after appropriate normalization. However, this is a heavily biased estimator because $\textrm{E}[X^2] \ne \textrm{E}[X]^2$. Instead we can use the replica trick\cite{Zhang1993,Hastings2010,Blunt2014,Overy2014} to estimate $(\sum_{i\in\mathcal{V}} H_{ai} c_i)^2$. Here, we perform two statistically independent FCIQMC simulations simultaneously, such that the two estimates of $-\Delta \tau \sum_{i\in\mathcal{V}} H_{ai} c_i$, labelled $S_a^1$ and $S_a^2$, are uncorrelated, and $\textrm{E}[S^1_a S^2_a] = \textrm{E}[S^1_a]\textrm{E}[S^2_a]$. Finally, by comparison between Eq.~(\ref{eq:E_2}) and Eq.~(\ref{eq:spawn_vec}), a stochastic estimate of $\Delta E_2$ at imaginary time $\tau$ may be constructed as
\begin{equation}
\Delta E_2(\tau) = \frac{1}{(\Delta\tau)^2} \sum_{a\notin\mathcal{V}} \frac{ S^1_a(\tau) S^2_a(\tau) }{ E_0 - H_{aa} },
\end{equation}
which should be normalized by $\bra \Psi^1(\tau) | \Psi^2(\tau) \ket$ (usually averaged separately, to avoid biases).

The zeroth-order energy appearing in the denominator, $E_0$, will also need to be sampled from FCIQMC, and so will be a random variable. One may then worry about a theoretical bias, as $\textrm{E}[X/Y] \ne \textrm{E}[X]/\textrm{E}[Y]$, which cannot be resolved through a replica trick. However, this bias should be small provided that the denominator is large compared to its stochastic noise. Because low-energy determinants are likely to be included in $\mathcal{V}$, $E_0$ is likely to be well separated to all $H_{aa}$. We present complete active space calculations in supplementary material, which demonstrate the very-high accuracy of this estimator.

Besides this theoretical bias, the expectation value of the above estimator will rigorously return the correct EN2 energy as from a non-QMC method. However, the above contribution cannot necessarily be sampled inexpensively; for many $\mathcal{V}$, efficient excitation generators are feasible that never create contributions outside this subspace. Allowing spawns outside $\mathcal{V}$ to calculate $S_a$ would then be an additional cost.

We now instead consider initiator FCIQMC, which has proven particularly accurate for a given number of simultaneously occupied determinants. As such, being able to perform an accurate perturbative correction beyond i-FCIQMC would be particularly powerful. Moreover, in the case of i-FCIQMC, calculation of $S_a = -\Delta \tau \sum_{i\in\mathcal{V}} H_{ai} c_i$ is already required to enforce the initiator criteria, and so accumulation of $\Delta E_2(\tau)$ is truly a small cost for a replica i-FCIQMC calculation.

As above, we intuitively would like to define $\mathcal{V}$ as the space in which projection is not truncated. However, the situation here is less clear than that above. Spawnings to occupied determinants in i-FCIQMC are always accepted, so all occupied determinants must lie within $\mathcal{V}$. But whether or not spawnings are truncated on an unoccupied determinant depends on the random number generator (RNG) state. Depending on this RNG state, a determinant may be spawned to once by a non-initiator (rejected), once by an initiator (accepted), or twice or more (accepted). As such, a fixed zeroth-order space cannot be defined. Instead, we consider the truncation to be dynamic, $\mathcal{V}(\tau)$, so that an EN2 correction may be calculated appropriately for each iteration, $\Delta E_2(\tau)$. This is done in the intuitive way: whenever a spawning is removed due to the initiator criteria, this must be viewed as a truncation, and so a contribution should be added to $\Delta E_2(\tau)$. In practice, a non-zero contribution requires a removal on both replicas.

However, if $\mathcal{V}(\tau)$ is defined as non-constant, then the FCIQMC wave function will not exactly sample the zeroth-order wave function in $\mathcal{V}(\tau)$ in the limit of large $\tau$. Nonetheless, if the effective truncated space varies only slowly (and only in regions where $|\Psi_0(\tau)\ket$ is small), then it is reasonable to assume that the expectation value of the FCIQMC wave function will be a good approximation to the true zeroth-order wave function. This approximation does not invalidate the approach - one can imagine starting from the exact $|\Psi_0(\tau)\ket$ and then varying this zeroth-order wave function. If the variation is small, then the change in $E_0$ and $\Delta E_2$ will also be small, and so an accurate $E_0 + \Delta E_2$ may be obtained. Ultimately, this accuracy can only be assessed by testing.

Another issue then arises, that there are multiple definitions of $E_0$ available for an inexact wave function. For an exact zeroth-order wave function, the energy estimator $ \bra \Psi_{\textrm{T}} | \hat{H}_0 | \Psi_0 \ket / \bra \Psi_{\textrm{T}} | \Psi_0 \ket $ is independent of the choice of non-orthogonal trial wave function, $| \Psi_{\textrm{T}} \ket$, which is indeed what we find for a constant truncation within FCIQMC. For an approximate $ | \Psi_0 \ket $, however, essentially any energy may be obtained depending on $| \Psi_{\textrm{T}} \ket$. For the combination of $E_0 + \Delta E_2$ to remain accurate, we want the most accurate estimate of the zeroth-order energy available. For this, we believe that the correct choice is the variational energy estimator
\begin{equation}
E_{\textrm{var}}(\tau) = \frac{ \bra \Psi^1(\tau) | \hat{H} | \Psi^2(\tau) \ket }{ \bra \Psi^1(\tau) | \Psi^2(\tau) \ket }.
\label{eq:var_est}
\end{equation}
Defining the FCIQMC wave function at $\tau$ as the exact zeroth-order wave function plus a correction, $ | \Psi(\tau) \ket = | \Psi_0(\tau) \ket + | \delta \Psi (\tau) \ket $ (and rescaling $ | \Psi_0(\tau) \ket $ so that $ \bra \Psi_0(\tau) | \delta \Psi (\tau) \ket = 0$), it can be seen that
\begin{equation}
E_{\textrm{var}}(\tau) = E_0(\tau) + \mathcal{O}(\delta \Psi^2).
\end{equation}
Meanwhile, the commonly-used FCIQMC estimators that project against a trial wave function (typically the Hartree--Fock determinant) have an error of $\mathcal{O}(\delta \Psi)$, and so for a small $| \delta \Psi(\tau) \ket$ will be less accurate, often significantly so. Note that $E_{\textrm{var}}$ is not necessarily more accurate as an estimate of the true ground-state energy. However, as a zeroth-order energy about which to add $\Delta E_2$, only $E_{\textrm{var}}$ is sensible.

To summarize the procedure:
\begin{enumerate}
\item Perform i-FCIQMC with two independent replica simulations, sampling $|\Psi^1(\tau)\ket$ and $|\Psi^2(\tau)\ket§$, and accumulating $ \bra \Psi^1(\tau) | \hat{H} | \Psi^2(\tau) \ket $ and $ \bra \Psi^1(\tau) | \Psi^2(\tau) \ket $.
\item Each iteration, for determinants $ | D_a \ket$ where spawnings are removed on \emph{both} simulations due to initiator criteria, label the removed contributions as $ S^1_a(\tau)$ and $ S^2_a(\tau)$. Then the contribution to $\Delta E_2$ from this iteration is
\begin{equation}
\Delta E_2(\tau) = \frac{1}{(\Delta\tau)^2} \sum_{a} \frac{ S_a^1(\tau) S_a^2(\tau) }{ E_{\textrm{var}}(\tau) - H_{aa} }.
\end{equation}
\item At the end of the simulation, the corrected energy is given by
\begin{equation}
E_{\textrm{tot}} = \frac{\textrm{E}[\; \bra \Psi^1(\tau) | \hat{H} | \Psi^2(\tau) \ket + \Delta E_2(\tau) \;]}{\textrm{E}[\; \bra \Psi^1(\tau) | \Psi^2(\tau) \ket \;]},
\end{equation}
with each expectation value $\textrm{E}[\; ... \;]$ estimated by an average over the simulation after convergence.
\end{enumerate}
This is trivial to implement in an existing FCIQMC code, provided that the $E_{\textrm{var}}$ estimator is available. The EN2 correction may be calculated in the excited-state FCIQMC algorithm\cite{Blunt2015_3} by exactly the same approach.

\begin{figure*}[t]
\includegraphics{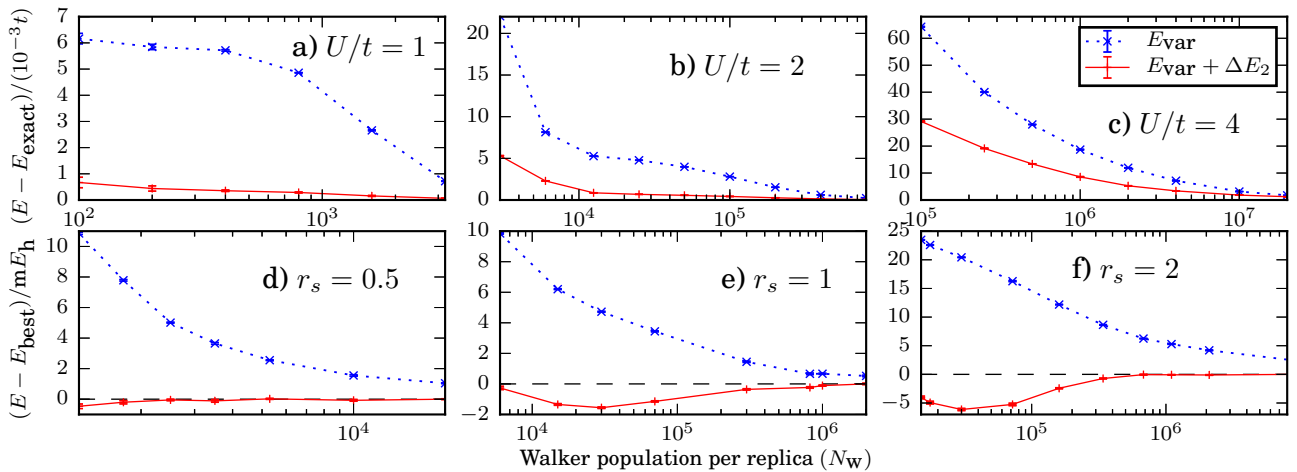}
\caption{Results for the periodic two-dimensional $18$-site Hubbard model at half-filling (a-c), and the 14-electron spin-unpolarized UEG in three dimensions, with 358 plane-wave spin orbitals (d-f), varying coupling strength ($U/t$ or $r_s$) and $\Nw$. For (d-f), dashed lines show near-exact best estimates.}
\label{fig:hub_ueg}
\end{figure*}
We now discuss the computational cost of this approach. The only additional cost for accumulating $\Delta E_2$, compared to basic i-FCIQMC, is in calculating each $H_{aa}$, which is essentially negligible compared to the rest of the simulation. A replica simulation must also be performed, doubling iteration time but also doubling the samples obtained, such that efficiency is unaffected; there is also a doubling of memory, but FCIQMC is significantly more time-limited than memory-limited. There is however a larger cost in calculating the variational energy estimate, $E_{\textrm{var}}$. In our current implementation this is done by first accumulating the FCIQMC two-body reduced density matrix (2-RDM), $\hat{\Gamma}$, as described in Refs.~(\onlinecite{Overy2014}) and (\onlinecite{Blunt2017}), with the numerator of the variational energy then obtained as $ \textrm{Tr} \big[ \hat{\Gamma} \hat{H} \big] $. Even with the many schemes described previously\cite{Overy2014,Blunt2017}, we currently find that accumulating 2-RDMs increases iteration time by a factor of $\sim 1.5-3$. Such RDM calculations are becoming the norm in FCIQMC, and the cost of calculating $E_{\textrm{var}}$ can likely be reduced, but for now we note this additional cost compared to a non-RDM simulation. Another concern is that $\Delta E_2$ may have larger noise than $E_{\textrm{var}}$, and so require additional sampling. Noise in $\Delta E_2$ is often larger than in $E_{\textrm{var}}$, but decreases more quickly with walker population, $\Nw$, and we often find that our usual protocol automatically gives sufficiently small error bars in both. However, for some challenging cases far from convergence, additional sampling may be required. Nonetheless, this additional sampling is far cheaper than instead reducing initiator error by increasing $\Nw$ without $\Delta E_2$.

\emph{Results:-} For results, FCIQMC simulations were performed using NECI\cite{NECI_github}, and SHCI benchmarks obtained using Dice, with integrals generated using PySCF\cite{pyscf}. All calculations use time-reversal symmetrized functions\cite{Smeyers1973} rather than Slater determinants.

We first apply the correction to the Hubbard model and uniform electron gas (UEG) at a range of coupling strengths. For the Hubbard model we study a periodic two-dimensional 18-site lattice at half-filling, using crystal momentum symmetry ($\boldsymbol{K}=\boldsymbol{0}$), at $U/t=1$, $2$ and $4$. For the UEG, we take the three-dimensional spin-unpolarized 14-electron system in a basis of 358 plane-wave spin orbitals, again restricting to $\boldsymbol{K}=\boldsymbol{0}$, and dimensionless density parameters ($r_s$) of $0.5$, $1.0$ and $2.0$.

Results are presented in Figure~\ref{fig:hub_ueg} (with data in supplementary material). For the Hubbard model at $U/t=1$ the correction is highly accurate, removing $>90-95\%$ of initiator error for all walker populations considered. With $100$ walkers, for example, initiator error is reduced from $6.2 \times 10^{-3}t$ to $0.7 \times 10^{-3} t$. At $U/t=2$ the correction is $\sim 75-85\%$. As expected, the correction is less effective at intermediate coupling, though still reducing initiator error from $64 \times 10^{-3} t$ to $29 \times 10^{-3} t$ with $10^5$ walkers.

\begin{table*}[t]
\begin{center}
{\footnotesize
\begin{tabular}{@{\extracolsep{4pt}}cc|c|ccccc@{}}
\hline                                    
\hline                                    
State & $R$/\AA & Benchmark/$\Eh$ & $E_{\textrm{var}}$/$\Eh$ & $(E_{\textrm{var}} + \Delta E_2)$/$\Eh$ & Initiator error/$\textrm{m}\Eh$ & Final error/$\textrm{m}{\Eh}$ & \% corrected \\
\hline                                    
$X{}^1\Sigma_g^{{}^+}$  & 1.24253 & -75.80271(2) & -75.80071(5) & -75.80272(7) & 1.93(5) & -0.01(7) & 101(3) \\
                        & 2.0     & -75.64565(2) & -75.64336(4) & -75.6452(1)  & 2.29(5)  & 0.4(1)   & 82(5) \\
\hline                                                                         
$B'{}^1\Sigma_g^{{}^+}$ & 1.24253 & -75.71213(2) & -75.71020(9) & -75.7121(1)  & 1.80(9) &  0.0(1)  & 98(6) \\
                        & 2.0     & -75.61486(2) & -75.61257(3) & -75.6145(1)  & 2.29(4) &  0.4(1)  & 82(5) \\
\hline                                    
\hline                                    
\end{tabular}
}
\caption{Energies for C$_2$ in a cc-pVQZ basis set, with 4 core electrons uncorrelated and at equilibrium and stretched geometries. $B'{}^1\Sigma_g^{{}^+}$ results use the excited-state FCIQMC algorithm\cite{Blunt2015_3}. Benchmarks are taken from Ref.~(\onlinecite{Holmes2017}), with associated errors marking the stated upper uncertainty. All FCIQMC calculations use $2 \times 10^5$ walkers per replica and state.}
\label{tab:c2_vqz}
\vspace{5mm}
{\footnotesize
\begin{tabular}{@{\extracolsep{4pt}}l|l|l|ccccc@{}}
\hline                                    
\hline                                    
System & Benchmark/$\Eh$ & $\Nw$ & $E_{\textrm{var}}$/$\Eh$ & $(E_{\textrm{var}} + \Delta E_2)$/$\Eh$ & Initiator error/$\textrm{m}\Eh$ & \% corrected \\
\hline                                    
Water (aug-cc-pVDZ)        & -76.274457(9) & $1 \times 10^3$ & -76.25901(3)  & -76.2750(1)   & 15.44(3) & 104(1) \\
Ethylene (cc-pVDZ)         & -78.35624(2)  & $1 \times 10^5$ & -78.35364(2)  & -78.3565(1)   & 2.60(3)  & 110(5) \\
Formaldehyde (aug-cc-pVDZ) & -114.2463(1)  & $5 \times 10^5$ & -114.24155(2) & -114.2461(2)  & 4.8(1)   & 96(4)  \\
Butadiene (ANO-L-pVDZ)     & -155.5582(1)  & $5 \times 10^7$ & -155.54323(8) & -155.5578(10) & 15.0(1)  & 98(7)  \\
\hline                                    
\hline                                    
\end{tabular}
}
\caption{Results for molecular systems at the given walker populations, $\Nw$, chosen so that significant initiator error exists. $1s$ cores were frozen in each case. The benchmark for butadiene is a SHCI result from Ref.~\onlinecite{Chien2018}. Other benchmarks were obtained from new extrapolated SHCI calculations. Results for additional $\Nw$ values are given in supplementary material.}
\label{tab:mol}
\end{center}
\end{table*}

For the UEG at $r_s=0.5$, results with the EN2 correction are always correct within $0.5$m$\Eh$, and always within $0.1$m$\Eh$ for $\Nw \ge 2500$, despite a Hilbert space dimension of $\sim10^{20}$. For comparison, initiator error is $\approx 5$m$\Eh$ at $\Nw=2500$. Although it is not surprising to find a perturbative correction to be effective at low $r_s$, this accuracy should address concerns about the validity of the correction within the initiator approximation. Large improvements are again made at $r_s=1.0$ and $2.0$, although decreasing somewhat in line with the increasing correlation strength. The results are non-variational here, but even at $r_s=2.0$ the correction is $\sim 110 - 130\%$.

The above UEG example should also address potential concerns about the use of replica sampling in large spaces with small walker populations. Despite a basis of $358$ spin orbitals and as little as $1.25 \times 10^3$ walkers, error bars in both $E_{\textrm{var}}$ and $\Delta E_2$ are well controlled.

In Table~\ref{tab:c2_vqz}, carbon dimer results are presented with a cc-pVQZ basis set, with $4$ core electrons uncorrelated, at equilibrium and stretched geometries. We use the excited-state FCIQMC algorithm\cite{Blunt2015_3} and study both the ground and first excited states of ${}^1\Sigma_g^{{}^+}$ character. This system has been previously studied with benchmark accuracy by the density matrix renormalization group algorithm (DMRG)\cite{Sharma2015}, FCIQMC\cite{Blunt2015_3} and SHCI\cite{Holmes2017}, and very recently by Sharma\cite{Sharma2018} and Guo \emph{et al.}\cite{Guo2018_2} in perturbative DMRG studies (although there correlating 12 electrons in a cc-pVDZ basis). In our previous FCIQMC study, it was found that $10^6$ walkers were required for an accuracy of $1$m$\Eh$ for both states and geometries. With $1.6 \times 10^7$ walkers, an accuracy of $\sim 0.1-0.2$m$\Eh$ was obtained.

Here we consider a much smaller $\Nw$ of $2\times 10^5$ to assess $\Delta E_2$. At equilibrium geometry, initiator error is around $1.8-1.9$m$\Eh$, which is removed effectively in its entirety (within stochastic errors of $0.1$m$\Eh$) by the correction. At the more strongly correlated $R=2.0$\AA$\,$ the correction is slightly less accurate, but still substantial, removing $82 \pm 5\%$ of initiator error. The correction is equally effective for the excited $B'{}^1\Sigma_g^{{}^+}$ state.

Results are presented in Table~\ref{tab:mol} for additional molecular systems, with geometries in supplementary material\cite{Hoy1979,Schreiber2008,Daday2012}. The most challenging system here is butadiene in an ANO-L-pVDZ basis, with $22$ electrons correlated in $82$ spatial orbitals, which was previously studied by i-FCIQMC\cite{Daday2012} using $10^9$ walkers. Subsequent extrapolated DMRG\cite{Olivares2015} and SHCI\cite{Chien2018} results agree that the ground-state energy from i-FCIQMC was too high by $\sim9$m$\Eh$, likely due to remaining initiator error, despite the large walker population. Here with a much smaller walker population of $5 \times 10^7$, almost all initiator error is removed by $\Delta E_2$. We note that the error bar on the corrected result is $1$m$\Eh$, so it is possible that the true agreement with SHCI is not as accurate as presented. However, even in the event that the quoted value is incorrect by $2$ standard errors, the EN2 correction still represents a dramatic improvement. These calculations did not require careful choice of molecular orbitals, which were always restricted Hartree--Fock orbitals. The computational resources for this study are modest compared to large-scale FCIQMC, using at most $320$ processor cores, while FCIQMC scales efficiently up to at least $10^4$ cores, a powerful possibility in combination with the correction presented here.

\emph{Conclusion:-} We have introduced the calculation of an EN2 correction from discarded spawning attempts in replica i-FCIQMC. The EN2 correction itself is essentially free to accumulate, although some additional cost may be required to accumulate the variational zeroth-order energy estimate, or to perform additional sampling. In non-strongly-correlated cases the correction regularly removes $>90\%$ of initiator error, and even in strongly-correlated regimes represents an important improvement. This correction therefore significantly extends the reach of FCIQMC, which we expect will be a powerful possibility in future applications of the method.

\emph{Supplementary Material:-} See supplementary material for data from Figure~\ref{fig:hub_ueg}, additional results for molecular systems, and discussion of theoretical biases.

\begin{acknowledgments}
We thank George Booth and Ali Alavi for helpful comments on this manuscript, and Werner Dobrautz and Olle Gunnarsson for providing 18-site Hubbard model FCI values. We are very grateful to St John's College, Cambridge for funding and supporting this work through a Research Fellowship. This study made use of the CSD3 Peta4 CPU cluster.
\end{acknowledgments}

%

\clearpage
\onecolumngrid
\begin{center}
\textbf{\large Supplementary material for ``An efficient and accurate perturbative correction to initiator full configuration interaction quantum Monte Carlo''}
\vspace{5mm}

Nick~S.~Blunt

\textit{University Chemical Laboratory, Lensfield Road, Cambridge, CB2 1EW, United Kingdom}
\vspace{8mm}
\end{center}
\setcounter{equation}{0}
\setcounter{figure}{0}
\setcounter{table}{0}
\setcounter{page}{1}
\makeatletter
\renewcommand{\theequation}{S\arabic{equation}}
\renewcommand{\thefigure}{S\arabic{figure}}
\renewcommand{\bibnumfmt}[1]{[S#1]}
\renewcommand{\citenumfont}[1]{S#1}

\twocolumngrid

\section{Application to traditional truncated spaces}

\begin{table*}[t]
\begin{center}
{\footnotesize
\begin{tabular}{@{\extracolsep{4pt}}ccccccc@{}}
\multicolumn{7}{l}{\large Ne aug-cc-pVDZ - CAS + EN2} \vspace{3mm} \\
\hline
\hline
 & \multicolumn{3}{c}{Wtih an exact $|\Psi_0\ket$} & \multicolumn{3}{c}{With a stochastic $|\Psi_0\ket$} \\
\cline{2-4} \cline{5-7}
Active space & $E_{\textrm{CAS}}/\Eh$ & $(E_{\textrm{CAS}} + \Delta E_2)/\Eh$ & $ \Delta E_2/\Eh$ & $E_{\textrm{CAS}}/\Eh$ & $(E_{\textrm{CAS}} + \Delta E_2)/\Eh$ & $ \Delta E_2/\Eh$ \\
\hline
     (8\El,8\Or)  &   -128.5026265  &  -128.74429(4)  &   -0.24167(4)  &   -128.502625(1)   &    -128.74430(4)    &    -0.24168(4)   \\
     (8\El,13\Or) &   -128.5294242  &  -128.73796(5)  &   -0.20854(5)  &   -128.529427(3)   &    -128.73798(5)    &    -0.20855(5)   \\
     (8\El,16\Or) &   -128.6131153  &  -128.71468(2)  &   -0.10156(2)  &   -128.61312(1)    &    -128.71469(1)    &    -0.10157(1)   \\
     (8\El,17\Or) &   -128.6428048  &  -128.71299(1)  &   -0.07018(1)  &   -128.64280(1)    &    -128.71299(2)    &    -0.07019(2)   \\
\hline
\hline
\end{tabular}
}
\caption{Comparison of CAS calculations wtih EN2 corrections, using FCIQMC as the CAS solver, both with an exact and stochastically-sampled zeroth-order wave function and energy. The system is Ne in an aug-cc-pVDZ basis set with $2$ core electrons uncorrelated, with an FCI energy of $-128.7094755\Eh$. Simulations with an exact zeroth-order wave function use semi-stochastic FCIQMC, but with the deterministic space set to $\mathcal{V}$, such that $E_{\textrm{CAS}}$ is obtained exactly.}
\label{tab:ne_cas}
\end{center}
\end{table*}

As discussed in the main text, as well as correcting initiator error, an FCIQMC-based estimate of $\Delta E_2$ may also be applied to standard truncated spaces, such as a complete active space (CAS). This allows a study of the EN2 correction in isolation, without additional complications arising from the unconventional nature of the initiator approximation.

We study such an example here, using FCIQMC to perform CAS calculations (using restricted Hartree--Fock (RHF) orbitals). Instead of removing the closed-shell and virtual orbitals from the simulation, we include them and allow spawning to basis states in which they have non-zero occupation. These spawnings to outside the CAS are then used to construct $\Delta E_2$, as described in the main text, and removed from the simulation afterwards.

We study the Ne atom in an aug-cc-pVDZ basis with $2$ uncorrelated core electrons, for which exact FCI may be performed. As well as performing stochastic FCIQMC within the CAS, we also perform a simulation with deterministic application of $\hat{P}$ within $\mathcal{V}$, allowing $|\Psi_{\textrm{CAS}}\ket$ and $E_{\textrm{CAS}}$ (the zeroth-order wave function and energy) to be obtained exactly. This is done using the semi-stochastic algorithm described in Refs.~(\onlinecite{S_Petruzielo2012}) and (\onlinecite{S_Blunt2015}), but setting the deterministic space to be the entirety of $\mathcal{V}$. Only spawning events outside of $\mathcal{V}$ are allowed, in order to construct $\Delta E_2$. In this way, the estimator for $\Delta E_2$ has no theoretical bias due to a stochastic zeroth-order energy estimate. This simulation is then compared to one without the semi-stochastic adaptation, where the zeroth-order wave function and energy are both obtained fully stochastically.

Results are presented in Table~\ref{tab:ne_cas}, for active spaces ranging from $(8\El,8\Or)$ to $(8\El,17\Or)$ (the aug-cc-pVDZ basis has $22$ spatial orbitals in total, after removing the $1s$ core). The results for both the zeroth-order energy estimate and the EN2 correction are identical between simulations with either a stochastic or exact $|\Psi_{\textrm{CAS}}\ket$, within stochastic error bars of $\sim 10^{-5}\Eh$, demonstrating the efficacy of the approach.

This also allows a comparison of noise in $\Delta E_2$ between simulations with different amounts of noise in the zeroth-order estimates. Results with active spaces of $(8\El,8\Or)$ and $(8\El,13\Or)$ were averaged over exactly $9 \times 10^5$ iterations, for both stochastic and deterministic cases, allowing direct comparison. For both $(8\El,8\Or)$ and $(8\El,13\Or)$ active spaces, error bars on estimates of $\Delta E_2$ are identical to $1$ significant figure. As such, the use of semi-stochastic within the zeroth-order space does not significantly impact the noise on $\Delta E_2$. Studying error bars to a second significant figure shows that results with a deterministic $|\Psi_{\textrm{CAS}}\ket$ do result in smaller noise on $\Delta E_2$, but not substantially so. This is not unreasonable - the number of contributions to $\Delta E_2$ is proportional to the square of the number of spawnings outside of $\mathcal{V}$. This density determines the noise on the $\Delta E_2$ estimate to a greater extent than the noise on the $| \Psi_0 \ket$ estimate.

Likewise, contributions to $\Delta E_2$ are not directly correlated from one iteration to the next, but only indirectly through the correlation between $\Psi_0(\tau)$ and $\Psi_0(\tau + \Delta \tau)$. We find that long autocorrelation lengths are not a significant issue for the sampling of $\Delta E_2$, compared to $E_0$.

\section{Data for model systems}

We include numerical data for the results of Figure~$1$ in the main text, studying the Hubbard model and the uniform electron gas (UEG) at a range of coupling strengths.

For the Hubbard model, an $18$-site periodic two-dimensional lattice is studied at $U/t=1$, $2$ and $4$. The Hilbert space dimension is $\sim 10^8$. The time step was kept at a constant value of $\Delta \tau=0.01$ and the initiator threshold was set to $n_a=3$. The semi-stochastic adaptation was used at $U/t=4$, with a deterministic space of dimension $10^4$, although this is not necessary and does not significantly change either initiator error or $\Delta E_2$.

For the UEG, a three-dimensional spin-unpolarized $14$-electron system is studied in a basis of 358 plane-wave spin orbitals, at $r_s=0.5$, $1$ and $2$. The Hilbert space dimension here is $\sim 10^{20}$. For $r_s=0.5$ a time step of $\Delta \tau = 4 \times 10^{-4}$ was used. At $r_s=1$ and $2$, the time step was varied to prevent creation of large spawning events, resulting in larger time steps of between $\Delta \tau = 1 \times 10^{-3}$ and $2 \times 10^{-3}$, although once again this difference does not significantly alter results, and the EN2 correction is well-behaved regardless.

All calculations use time-reversal symmetrized functions\cite{S_Smeyers1973} rather than Slater determinants.

\section{Additional data}

We present further data for molecular systems at a range of walker populations, $\Nw$. The molecules studied are C$_2$ in a cc-pVTZ basis (both ground and first excited states of ${}^1\Sigma_g^{{}^+}$ character, at equilibrium and stretched geometries), water in an aug-cc-pVDZ basis, ethylene in a cc-pVDZ basis and formaldehyde in an aug-cc-pVDZ basis. $1s$ cores are uncorrelated in each case.

For water, the semi-stochastic adaptation was not used. For C2, ethylene and formaldehyde, semi-stochastic was used with a deterministic space of dimension $10^4$, by picking the most populated basis states upon convergence, as described in Ref.~(\onlinecite{S_Blunt2015}). For butadiene (results presented in the main text), a deterministic space of dimension $2 \times 10^5$ was used, chosen using the same scheme. Results were typically averaged over between $5 \times 10^4$ and $10^6$ iterations. The initiator threshold was taken as $n_a=3$, and the time step was varied in the early stages of the simulation to prevent bloom events (defined as a single spawning event with magnitude greater than $n_a$).

Benchmarks for C$_2$ cc-pVTZ are taken from Ref.~(\onlinecite{S_Blunt2015_3}). Benchmarks for other systems are obtained from semistochastic heat-bath configuration interaction (SHCI) calculations\cite{S_Holmes2016_2, S_Sharma2017} performed with Dice, with quadratic extrapolations as described in Ref.~(\onlinecite{S_Chien2018}) (although using RHF orbitals). Time-reversal symmetrized functions\cite{S_Smeyers1973} are used for both FCIQMC and SHCI. For formaldehyde, the smallest threshold for the variational SHCI step was $\epsilon_{\textrm{V}} = 10^{-5}$, corresponding to $\sim 3.7 \times 10^7$ basis states in $\mathcal{V}$. The perturbative threshold was set to $\epsilon_{\textrm{PT}}=10^{-10}$ in all cases.

The geometry for water is from Ref.~(\onlinecite{S_Hoy1979}), the geometry for ethylene is taken from Ref.~(\onlinecite{S_Schreiber2008}), and the geometry for butadiene is from Ref.~(\onlinecite{S_Daday2012}).

\makeatletter\onecolumngrid@push\makeatother
\begin{table*}
\begin{center}
{\footnotesize
\begin{tabular}{@{\extracolsep{4pt}}ccccccc@{}}
\multicolumn{7}{l}{\large $18$-site Hubbard model} \vspace{3mm} \\
\hline
\hline
$\Nw$ & $E_{\textrm{var}}/t$ & $(E_{\textrm{var}} + \Delta E_2)/t$ & $\Delta E_2/(10^{-3} t)$ & Initiator error/$(10^{-3} t)$ & Final error/$(10^{-3} t)$ & \% corrected \\ 
\hline
      100  &    -27.6886(2)  &    -27.6941(2)  &       -5.49(7)  &         6.2(2)  &         0.7(2)  &          89(3)  \\
      200  &    -27.6889(1)  &    -27.6943(1)  &       -5.40(5)  &         5.8(1)  &         0.4(1)  &          92(2)  \\
      400  &   -27.68900(3)  &   -27.69437(3)  &       -5.36(2)  &        5.72(3)  &        0.35(3)  &        93.8(6)  \\
      800  &   -27.68986(1)  &   -27.69443(2)  &      -4.572(8)  &        4.86(1)  &        0.29(2)  &        94.1(3)  \\
     1600  &   -27.69206(2)  &   -27.69456(2)  &      -2.504(8)  &        2.66(2)  &        0.16(2)  &        94.1(8)  \\
     3200  &   -27.69400(1)  &   -27.69466(1)  &      -0.654(5)  &        0.72(1)  &        0.06(1)  &          91(1)  \\
\hline
\multicolumn{2}{l}{Exact energy/$t$} &   -27.69472  & & & & \\ 
\hline
\hline
\end{tabular}
}
\caption{Total energies for the periodic two-dimensional $18$-site Hubbard model at half-filling, at $U/t=1$.}
{\footnotesize
\vspace{1cm}
\begin{tabular}{@{\extracolsep{4pt}}ccccccc@{}}
\hline
\hline
$\Nw$ & $E_{\textrm{var}}/t$ & $(E_{\textrm{var}} + \Delta E_2)/t$ & $\Delta E_2/(10^{-3} t)$ & Initiator error/$(10^{-3} t)$ & Final error/$(10^{-3} t)$ & \% corrected \\ 
\hline
   $1.5\times10^3$ &   -23.74375(5)  &   -23.77703(6)  &      -33.28(3)  &       41.65(5)  &        8.37(6)  &        79.9(1)  \\
   $3\times10^3$   &   -23.76334(3)  &   -23.78009(4)  &      -16.75(2)  &       22.06(3)  &        5.31(4)  &        75.9(1)  \\
   $6\times10^3$   &   -23.77726(2)  &   -23.78311(2)  &       -5.85(1)  &        8.14(2)  &        2.29(2)  &        71.9(2)  \\
 $1.25\times10^4$  &   -23.78013(1)  &   -23.78454(1)  &      -4.412(5)  &        5.27(1)  &        0.86(1)  &        83.7(2)  \\
 $2.5\times10^4$   &  -23.780632(8)  &  -23.784725(8)  &      -4.093(2)  &       4.771(8)  &       0.678(8)  &        85.8(2)  \\
   $5\times10^4$   &  -23.781424(7)  &  -23.784840(7)  &      -3.416(1)  &       3.979(7)  &       0.563(7)  &        85.9(2)  \\
   $1\times10^5$   &  -23.782599(7)  &  -23.784976(7)  &     -2.3768(7)  &       2.804(7)  &       0.427(7)  &        84.8(2)  \\
   $2\times10^5$   &  -23.783874(8)  &  -23.785137(8)  &     -1.2622(5)  &       1.528(8)  &       0.266(8)  &        82.6(4)  \\
   $4\times10^5$   &  -23.784787(9)  &  -23.785271(9)  &     -0.4845(4)  &       0.616(9)  &       0.132(9)  &          79(1)  \\
   $8\times10^5$   &  -23.785173(6)  &  -23.785354(6)  &     -0.1800(3)  &       0.229(6)  &       0.049(6)  &          78(2)  \\
\hline
\multicolumn{2}{l}{Exact energy/$t$} & -23.78540 & & & & \\ 
\hline
\hline
\end{tabular}
}
\caption{Total energies for the periodic two-dimensional $18$-site Hubbard model at half-filling, at $U/t=2$.}
{\footnotesize
\vspace{1cm}
\begin{tabular}{@{\extracolsep{4pt}}ccccccc@{}}
\hline
\hline
$\Nw$ & $E_{\textrm{var}}/t$ & $(E_{\textrm{var}} + \Delta E_2)/t$ & $\Delta E_2/(10^{-3} t)$ & Initiator error/$(10^{-3} t)$ & Final error/$(10^{-3} t)$ & \% corrected \\ 
\hline
  $1\times10^5$   &    -17.1881(2)  &    -17.2232(2)  &       -35.1(2)  &        64.2(2)  &        29.1(2)  &        54.6(3)  \\
  $2.5\times10^5$ &   -17.21235(8)  &    -17.2332(1)  &      -20.89(6)  &       40.04(8)  &        19.1(1)  &        52.2(2)  \\
  $5\times10^5$   &   -17.22437(6)  &   -17.23901(7)  &      -14.63(3)  &       28.01(6)  &       13.38(7)  &        52.2(2)  \\
  $1\times10^6$   &    -17.2337(1)  &    -17.2438(1)  &      -10.12(5)  &        18.7(1)  &         8.6(2)  &        54.1(5)  \\
  $2\times10^6$   &   -17.24045(5)  &   -17.24712(5)  &       -6.67(2)  &       11.93(5)  &        5.27(5)  &        55.9(3)  \\
  $4\times10^6$   &   -17.24524(4)  &   -17.24900(4)  &       -3.76(1)  &        7.14(4)  &        3.38(4)  &        52.6(4)  \\
  $1\times10^7$   &   -17.24925(4)  &   -17.25058(4)  &      -1.322(7)  &        3.13(4)  &        1.81(4)  &        42.2(6)  \\
  $2\times10^7$   &   -17.25065(4)  &   -17.25116(4)  &      -0.515(4)  &        1.74(4)  &        1.22(4)  &        29.6(8)  \\
\hline
\multicolumn{2}{l}{Exact energy/$t$} & -17.25239 & & & & \\ 
\hline
\hline
\end{tabular}
}
\caption{Total energies for the periodic two-dimensional $18$-site Hubbard model at half-filling, at $U/t=4$.}
\end{center}
\end{table*}
\clearpage
\makeatletter\onecolumngrid@pop\makeatother

\makeatletter\onecolumngrid@push\makeatother
\begin{table*}
\begin{center}
{\footnotesize
\vspace{1cm}
\begin{tabular}{@{\extracolsep{4pt}}ccccccc@{}}
\multicolumn{7}{l}{\large Uniform electron gas} \vspace{3mm} \\
\hline
\hline
$\Nw$ & $E_{\textrm{var}}/\Eh$ & $(E_{\textrm{var}} + \Delta E_2)/\Eh$ & $\Delta E_2/\textrm{m}\Eh$ & Initiator error/$\textrm{m}\Eh$ & Final error/$\textrm{m}{\Eh}$ & \% corrected \\ 
\hline
     1250  &    -0.56906(7)  &     -0.5804(1)  &       -11.3(1)  &       10.85(7)  &        -0.5(2)  &         104(1)  \\
     1750  &    -0.57212(5)  &     -0.5801(1)  &       -7.98(10)  &        7.79(6)  &        -0.2(1)  &         103(1) \\
     2500  &    -0.57490(3)  &    -0.57996(5)  &       -5.06(4)  &        5.01(4)  &       -0.05(6)  &         101(1)  \\
     3500  &    -0.57626(6)  &    -0.58002(9)  &       -3.76(8)  &        3.65(7)  &       -0.11(10)  &         103(3) \\
     5300  &    -0.57736(3)  &    -0.57989(4)  &       -2.53(3)  &        2.55(4)  &        0.02(5)  &          99(2)  \\
    10000  &    -0.57836(5)  &    -0.57998(6)  &       -1.63(3)  &        1.55(6)  &       -0.07(7)  &         105(5)  \\
    20000  &    -0.57887(3)  &    -0.57991(3)  &       -1.04(1)  &           -     &           -     &           -     \\
\hline
\multicolumn{2}{l}{Best estimate/$\Eh$} &  -0.57991(3)  & & & & \\ 
\hline
\hline
\end{tabular}
}
\caption{Correlation energies for the 14-electron UEG in three dimensions, in a basis of 358 plane-wave spin orbitals, at $r_s=0.5$. The benchmark is taken from the largest $\Nw$ considered, although we note that an i-FCIQMC value of $-0.5798(3)\Eh$ from Ref.~(\onlinecite{S_Neufeld2017}) is within error bars.}
{\footnotesize
\vspace{1cm}
\begin{tabular}{@{\extracolsep{4pt}}ccccccc@{}}
\hline
\hline
$\Nw$ & $E_{\textrm{var}}/\Eh$ & $(E_{\textrm{var}} + \Delta E_2)/\Eh$ & $\Delta E_2/\textrm{m}\Eh$ & Initiator error/$\textrm{m}\Eh$ & Final error/$\textrm{m}{\Eh}$ & \% corrected \\
\hline
 $6 \times 10^3$   &    -0.50887(4)  &     -0.5190(1)  &       -10.1(1)  &        9.85(6)  &        -0.3(1)  &         103(1)  \\
 $1.5 \times 10^4$ &    -0.51252(3)  &    -0.52007(6)  &       -7.55(5)  &        6.20(6)  &       -1.34(8)  &         122(1)  \\
 $3 \times 10^4$   &    -0.51400(3)  &    -0.52028(5)  &       -6.28(4)  &        4.72(6)  &       -1.56(7)  &         133(2)  \\
 $7 \times 10^4$   &    -0.51528(3)  &    -0.51987(4)  &       -4.59(2)  &        3.44(6)  &       -1.14(6)  &         133(2)  \\
 $3 \times 10^5$   &    -0.51728(4)  &    -0.51908(5)  &       -1.80(1)  &        1.44(7)  &       -0.36(7)  &         125(6)  \\
 $8.2 \times 10^5$ &    -0.51806(5)  &    -0.51895(5)  &      -0.891(7)  &        0.66(7)  &       -0.23(7)  &         135(15)  \\
 $1 \times 10^6$   &    -0.51806(4)  &    -0.51883(4)  &      -0.776(7)  &        0.67(6)  &       -0.11(6)  &         116(11)  \\
 $2 \times 10^6$   &    -0.51820(5)  &    -0.51872(5)  &      -0.522(3)  &           -     &           -     &           -      \\
\hline
\multicolumn{2}{l}{Best estimate/$\Eh$} &  -0.51872(5)  & & & & \\ 
\hline
\hline
\end{tabular}
}
\caption{Correlation energies for the 14-electron UEG in three dimensions, in a basis of 358 plane-wave spin orbitals, at $r_s=1.0$. The benchmark is taken from the largest $\Nw$ considered, although we note an i-FCIQMC value of $-0.51880(2)\Eh$ from Ref.~(\onlinecite{S_Neufeld2017}), showing good agreement.}
{\footnotesize
\vspace{1cm}
\begin{tabular}{@{\extracolsep{4pt}}ccccccc@{}}
\hline
\hline
$\Nw$ & $E_{\textrm{var}}/\Eh$ & $(E_{\textrm{var}} + \Delta E_2)/\Eh$ & $\Delta E_2/\textrm{m}\Eh$ & Initiator error/$\textrm{m}\Eh$ & Final error/$\textrm{m}{\Eh}$ & \% corrected \\ 
\hline
 $1.5 \times 10^4$  &    -0.41211(5)  &     -0.4396(2)  &       -27.5(2)  &       23.48(7)  &        -4.0(2)  &         117(1)  \\
 $1.75 \times 10^4$ &    -0.41300(5)  &     -0.4405(2)  &       -27.5(2)  &       22.59(8)  &        -4.9(3)  &         122(1)  \\
 $3 \times 10^4$    &    -0.41515(5)  &     -0.4417(2)  &       -26.6(2)  &       20.43(8)  &        -6.1(2)  &         130(1)  \\
 $7.2 \times 10^4$  &    -0.41931(9)  &     -0.4408(2)  &       -21.5(2)  &        16.3(1)  &        -5.2(2)  &         132(1)  \\
 $1.6 \times 10^5$  &    -0.42340(5)  &    -0.43801(10)  &      -14.61(8)  &       12.18(8)  &        -2.4(1)  &        120(1)  \\
 $3.4 \times 10^5$  &    -0.42694(5)  &    -0.43631(7)  &       -9.36(6)  &        8.64(8)  &       -0.72(10)  &        108(1)  \\
 $6.8 \times 10^5$  &    -0.42934(5)  &    -0.43563(8)  &       -6.29(6)  &        6.25(8)  &       -0.05(10)  &        101(2)  \\
 $1.1 \times 10^6$  &    -0.43030(6)  &    -0.43566(5)  &       -5.36(3)  &        5.29(8)  &       -0.07(8)  &         101(2)  \\
 $2.1 \times 10^6$  &    -0.43138(6)  &    -0.43569(6)  &       -4.31(3)  &        4.20(9)  &       -0.11(9)  &         103(2)  \\
 $8.9 \times 10^6$  &    -0.43314(5)  &    -0.43558(6)  &       -2.45(1)  &           -     &           -     &           -     \\
\hline
\multicolumn{2}{l}{Best estimate/$\Eh$} & -0.435584(58)  & & & & \\ 
\hline
\hline
\end{tabular}
}
\caption{Correlation energies for the 14-electron UEG in three dimensions, in a basis of 358 plane-wave spin orbitals, at $r_s=2.0$. The benchmark is taken from the largest $\Nw$ considered, $\Nw = 8.9 \times 10^6$. We could not find independent i-FCIQMC results for comparison, although stochastic coupled cluster results for CCSD and CCSDT from Ref.~(\onlinecite{S_Neufeld2017}) are -0.40181(4)$\Eh$ and -0.43212(7)$\Eh$, respectively.}
\end{center}
\end{table*}
\clearpage
\makeatletter\onecolumngrid@pop\makeatother

\begin{table*}[t]
\begin{center}
{\footnotesize
\begin{tabular}{@{\extracolsep{4pt}}cccccccc@{}}
\multicolumn{8}{l}{\large C$_2$ cc-pVTZ} \vspace{3mm} \\
\hline
\hline
State & $R$/\AA & Best estimate/$\Eh$ & $E_{\textrm{var}}$/$\Eh$ & $(E_{\textrm{var}} + \Delta E_2)$/$\Eh$ & Initiator error/$\textrm{m}\Eh$ & Final error/$\textrm{m}{\Eh}$ & \% corrected \\
\hline                                    
$X{}^1\Sigma_g^{{}^+}$  & 1.25 & -75.78515(10) & -75.78382(1) & -75.78519(3) & 1.3(1) & 0.0(1) & 103(8) \\
                        & 2.0  & -75.63095(10) & -75.62939(3) & -75.63087(5) & 1.6(1) & 0.1(1) & 95(8)  \\
\hline                                                          
$B'{}^1\Sigma_g^{{}^+}$ & 1.25 & -75.69572(10) & -75.69455(2) & -75.69578(4) & 1.2(1) & 0.1(1) & 105(9) \\
                        & 2.0  & -75.60145(10) & -75.59996(3) & -75.60138(6) & 1.5(1) & 0.1(1) & 96(8)  \\
\hline
\hline
\end{tabular}
}
\caption{Energies for the carbon dimer in a cc-pVTZ basis set, with 4 core electrons uncorrelated, at equilibrium and stretched geometries. $B'{}^1\Sigma_g^{{}^+}$ results use the excited-state FCIQMC algorithm. Best estimates are taken from Ref.~(\onlinecite{S_Blunt2015_3}). All FCIQMC calculations use $1 \times 10^5$ walkers per replica and state.}
\vspace{1cm}
{\footnotesize
\begin{tabular}{@{\extracolsep{4pt}}ccccccc@{}}
\multicolumn{7}{l}{\large Water aug-cc-pVDZ} \vspace{3mm} \\
\hline
\hline
$\Nw$ & $E_{\textrm{var}}/\Eh$ & $(E_{\textrm{var}} + \Delta E_2)/\Eh$ & $\Delta E_2/\textrm{m}\Eh$ & Initiator error/$\textrm{m}\Eh$ & Final error/$\textrm{m}{\Eh}$ & \% corrected \\ 
\hline
 $1 \times 10^3$   &   -76.25901(3)  &    -76.2750(1)  &       -16.0(2)  &       15.44(3)  &        -0.6(2)  &         104(1) \\
 $1 \times 10^4$   &   -76.27024(2)  &   -76.27393(6)  &       -3.69(5)  &        4.22(2)  &        0.53(6)  &          87(1) \\
 $4 \times 10^4$   &   -76.27306(2)  &   -76.27439(4)  &       -1.33(4)  &        1.40(2)  &        0.07(4)  &          95(3) \\
 $8 \times 10^4$   &   -76.27357(2)  &   -76.27447(3)  &       -0.90(2)  &        0.89(2)  &       -0.01(4)  &         101(4) \\
 $1.6 \times 10^5$ &   -76.27383(2)  &   -76.27446(2)  &      -0.636(9)  &        0.63(2)  &       -0.01(2)  &         101(4) \\
\hline
\multicolumn{2}{l}{Benchmark energy/$\Eh$} &  -76.274457(9)  & & & & \\ 
\hline
\hline
\end{tabular}
}
\caption{Energies for the water molecule at equilibrium geometry in the aug-cc-pVDZ basis set. The benchmark energy is obtained from SHCI calculations with a (very small) extrapolation.}
\vspace{1cm}
{\footnotesize
\begin{tabular}{@{\extracolsep{4pt}}ccccccc@{}}
\multicolumn{7}{l}{\large Ethylene cc-pVDZ} \vspace{3mm} \\
\hline
\hline
$\Nw$ & $E_{\textrm{var}}/\Eh$ & $(E_{\textrm{var}} + \Delta E_2)/\Eh$ & $\Delta E_2/\textrm{m}\Eh$ & Initiator error/$\textrm{m}\Eh$ & Final error/$\textrm{m}{\Eh}$ & \% corrected \\ 
\hline
 $1 \times 10^5$   &   -78.35364(2)  &    -78.3565(1)  &        -2.8(1)  &        2.60(3)  &        -0.2(1)  &         110(5)  \\
 $2.5 \times 10^5$ &   -78.35409(2)  &   -78.35639(6)  &       -2.30(6)  &        2.15(3)  &       -0.15(6)  &         107(3)  \\
 $5 \times 10^5$   &   -78.35440(1)  &   -78.35635(3)  &       -1.95(3)  &        1.84(2)  &       -0.11(3)  &         106(2)  \\
 $5 \times 10^6$   &   -78.35561(1)  &   -78.35622(1)  &      -0.615(5)  &        0.63(2)  &        0.02(2)  &          97(4)  \\
 $1 \times 10^7$   &  -78.355832(7)  &  -78.356227(6)  &      -0.394(3)  &        0.41(2)  &        0.01(2)  &          97(5)  \\
\hline
\multicolumn{2}{l}{Benchmark energy/$\Eh$} &   -78.35624(2)  & & & & \\ 
\hline
\hline
\end{tabular}
}
\caption{Energies for the ethylene molecule at equilibrium geometry in the cc-pVDZ basis set. The benchmark energy is obtained from SHCI calculations with a (very small) extrapolation.}
\vspace{1cm}
{\footnotesize
\begin{tabular}{@{\extracolsep{4pt}}ccccccc@{}}
\multicolumn{7}{l}{\large Formaldehyde aug-cc-pVDZ} \vspace{3mm} \\
\hline
\hline
$\Nw$ & $E_{\textrm{var}}/\Eh$ & $(E_{\textrm{var}} + \Delta E_2)/\Eh$ & $\Delta E_2/\textrm{m}\Eh$ & Initiator error/$\textrm{m}\Eh$ & Final error/$\textrm{m}{\Eh}$ & \% corrected \\ 
\hline
 $5 \times 10^5$  &  -114.24155(2)  &   -114.2461(2)  &        -4.6(2)  &         4.8(1)  &         0.2(2)  &          96(4)  \\
 $1 \times 10^6$  &  -114.24244(2)  &   -114.2462(1)  &        -3.8(1)  &         3.9(1)  &         0.1(2)  &          97(4)  \\
 $2 \times 10^6$  &  -114.24316(2)  &  -114.24604(7)  &       -2.88(7)  &         3.1(1)  &         0.3(1)  &          92(4)  \\
 $4 \times 10^6$  &  -114.24374(1)  &  -114.24612(3)  &       -2.37(3)  &         2.6(1)  &         0.2(1)  &          93(4)  \\
\hline
\multicolumn{2}{l}{Best estimate/$\Eh$} &  -114.2463(1)  & & & & \\ 
\hline
\hline
\end{tabular}
}
\caption{Energies for the formaldehyde molecule at equilibrium geometry in the aug-cc-pVDZ basis set. The benchmark energy is obtained from SHCI calculations with an extrapolation.}
\end{center}
\end{table*}

\begin{table}
{\large \textbf Geometries (\AA)}
\vspace{0.5cm}
\begin{flushleft}
{\setlength{\tabcolsep}{1em}
\texttt{
\vspace{1cm}
\begin{tabular}{@{}lrrr@{}}
\texttt{Water} & & & \vspace{5mm} \\
O & 0.0000 &  0.0000  &  0.1173 \\
H & 0.0000 &  0.7572  & -0.4692 \\
H & 0.0000 & -0.7572  & -0.4692 \\
\end{tabular}
\vspace{1cm}
\begin{tabular}{@{}lrrr@{}}
\texttt{Ethylene} & & & \vspace{5mm} \\
H & 0.000000 &  0.923274 &  1.238289 \\
H & 0.000000 & -0.923274 &  1.238289 \\
H & 0.000000 &  0.923274 & -1.238289 \\
H & 0.000000 & -0.923274 & -1.238289 \\
C & 0.000000 &  0.000000 &  0.668188 \\
C & 0.000000 &  0.000000 & -0.668188 \\
\end{tabular}
\vspace{1cm}
\begin{tabular}{@{}lrrr@{}}
\texttt{Formaldehyde} & & & \vspace{5mm} \\
O & 0.000000 &  0.0000 &  1.2050 \\
C & 0.000000 &  0.0000 &  0.0000 \\
H & 0.000000 &  0.9429 & -0.5876 \\
H & 0.000000 & -0.9429 & -0.5876 \\
\end{tabular}
\vspace{1cm}
\begin{tabular}{@{}lrrr@{}}
\texttt{Butadiene} & & & \vspace{5mm} \\
C & 0.000000 &  1.834350 & -0.157794 \\ 
C & 0.000000 & -1.834350 &  0.157794 \\
C & 0.000000 &  0.612753 &  0.388232 \\
C & 0.000000 & -0.612753 & -0.388232 \\
H & 0.000000 &  0.509700 &  1.466975 \\
H & 0.000000 & -0.509700 & -1.466975 \\
H & 0.000000 &  2.723649 &  0.452738 \\
H & 0.000000 & -2.723649 & -0.452738 \\
H & 0.000000 &  1.961466 & -1.231090 \\
H & 0.000000 & -1.961466 &  1.231090 \\
\end{tabular}
}}
\end{flushleft}
\end{table}


\begin{thebibliography}{54}%
\makeatletter
\providecommand \@ifxundefined [1]{%
 \@ifx{#1\undefined}
}%
\providecommand \@ifnum [1]{%
 \ifnum #1\expandafter \@firstoftwo
 \else \expandafter \@secondoftwo
 \fi
}%
\providecommand \@ifx [1]{%
 \ifx #1\expandafter \@firstoftwo
 \else \expandafter \@secondoftwo
 \fi
}%
\providecommand \natexlab [1]{#1}%
\providecommand \enquote  [1]{``#1''}%
\providecommand \bibnamefont  [1]{#1}%
\providecommand \bibfnamefont [1]{#1}%
\providecommand \citenamefont [1]{#1}%
\providecommand \href@noop [0]{\@secondoftwo}%
\providecommand \href [0]{\begingroup \@sanitize@url \@href}%
\providecommand \@href[1]{\@@startlink{#1}\@@href}%
\providecommand \@@href[1]{\endgroup#1\@@endlink}%
\providecommand \@sanitize@url [0]{\catcode `\\12\catcode `\$12\catcode
  `\&12\catcode `\#12\catcode `\^12\catcode `\_12\catcode `\%12\relax}%
\providecommand \@@startlink[1]{}%
\providecommand \@@endlink[0]{}%
\providecommand \url  [0]{\begingroup\@sanitize@url \@url }%
\providecommand \@url [1]{\endgroup\@href {#1}{\urlprefix }}%
\providecommand \urlprefix  [0]{URL }%
\providecommand \Eprint [0]{\href }%
\providecommand \doibase [0]{http://dx.doi.org/}%
\providecommand \selectlanguage [0]{\@gobble}%
\providecommand \bibinfo  [0]{\@secondoftwo}%
\providecommand \bibfield  [0]{\@secondoftwo}%
\providecommand \translation [1]{[#1]}%
\providecommand \BibitemOpen [0]{}%
\providecommand \bibitemStop [0]{}%
\providecommand \bibitemNoStop [0]{.\EOS\space}%
\providecommand \EOS [0]{\spacefactor3000\relax}%
\providecommand \BibitemShut  [1]{\csname bibitem#1\endcsname}%
\let\auto@bib@innerbib\@empty
\bibitem [{\citenamefont {Booth}, \citenamefont {Thom},\ and\ \citenamefont
  {Alavi}(2009)}]{Booth2009}%
  \BibitemOpen
  \bibfield  {author} {\bibinfo {author} {\bibfnamefont {G.~H.}\ \bibnamefont
  {Booth}}, \bibinfo {author} {\bibfnamefont {A.~J.~W.}\ \bibnamefont {Thom}},
  \ and\ \bibinfo {author} {\bibfnamefont {A.}~\bibnamefont {Alavi}},\
  }\href@noop {} {\bibfield  {journal} {\bibinfo  {journal} {J. Chem. Phys.}\
  }\textbf {\bibinfo {volume} {131}},\ \bibinfo {pages} {054106} (\bibinfo
  {year} {2009})}\BibitemShut {NoStop}%
\bibitem [{\citenamefont {Booth}\ \emph {et~al.}(2012)\citenamefont {Booth},
  \citenamefont {Gruneis}, \citenamefont {Kresse},\ and\ \citenamefont
  {Alavi}}]{Booth2012}%
  \BibitemOpen
  \bibfield  {author} {\bibinfo {author} {\bibfnamefont {G.~H.}\ \bibnamefont
  {Booth}}, \bibinfo {author} {\bibfnamefont {A.}~\bibnamefont {Gruneis}},
  \bibinfo {author} {\bibfnamefont {G.}~\bibnamefont {Kresse}}, \ and\ \bibinfo
  {author} {\bibfnamefont {A.}~\bibnamefont {Alavi}},\ }\href@noop {}
  {\bibfield  {journal} {\bibinfo  {journal} {Nature}\ }\textbf {\bibinfo
  {volume} {493}},\ \bibinfo {pages} {365} (\bibinfo {year}
  {2012})}\BibitemShut {NoStop}%
\bibitem [{\citenamefont {Shepherd}\ \emph {et~al.}(2012)\citenamefont
  {Shepherd}, \citenamefont {Gr\"{u}neis}, \citenamefont {Booth}, \citenamefont
  {Kresse},\ and\ \citenamefont {Alavi}}]{Shepherd2012_3}%
  \BibitemOpen
  \bibfield  {author} {\bibinfo {author} {\bibfnamefont {J.~J.}\ \bibnamefont
  {Shepherd}}, \bibinfo {author} {\bibfnamefont {A.}~\bibnamefont
  {Gr\"{u}neis}}, \bibinfo {author} {\bibfnamefont {G.~H.}\ \bibnamefont
  {Booth}}, \bibinfo {author} {\bibfnamefont {G.}~\bibnamefont {Kresse}}, \
  and\ \bibinfo {author} {\bibfnamefont {A.}~\bibnamefont {Alavi}},\
  }\href@noop {} {\bibfield  {journal} {\bibinfo  {journal} {Phys. Rev. B}\
  }\textbf {\bibinfo {volume} {86}},\ \bibinfo {pages} {035111} (\bibinfo
  {year} {2012})}\BibitemShut {NoStop}%
\bibitem [{\citenamefont {Thomas}, \citenamefont {Booth},\ and\ \citenamefont
  {Alavi}(2015)}]{Thomas2015}%
  \BibitemOpen
  \bibfield  {author} {\bibinfo {author} {\bibfnamefont {R.~E.}\ \bibnamefont
  {Thomas}}, \bibinfo {author} {\bibfnamefont {G.~H.}\ \bibnamefont {Booth}}, \
  and\ \bibinfo {author} {\bibfnamefont {A.}~\bibnamefont {Alavi}},\
  }\href@noop {} {\bibfield  {journal} {\bibinfo  {journal} {Phys. Rev. Lett.}\
  }\textbf {\bibinfo {volume} {114}},\ \bibinfo {pages} {033001} (\bibinfo
  {year} {2015})}\BibitemShut {NoStop}%
\bibitem [{\citenamefont {Thom}(2010)}]{Thom2012}%
  \BibitemOpen
  \bibfield  {author} {\bibinfo {author} {\bibfnamefont {A.~J.~W.}\
  \bibnamefont {Thom}},\ }\href@noop {} {\bibfield  {journal} {\bibinfo
  {journal} {Phys. Rev. Lett.}\ }\textbf {\bibinfo {volume} {105}},\ \bibinfo
  {pages} {263004} (\bibinfo {year} {2010})}\BibitemShut {NoStop}%
\bibitem [{\citenamefont {Scott}\ and\ \citenamefont {Thom}(2017)}]{Scott2017}%
  \BibitemOpen
  \bibfield  {author} {\bibinfo {author} {\bibfnamefont {C.~J.~C.}\
  \bibnamefont {Scott}}\ and\ \bibinfo {author} {\bibfnamefont {A.~J.~W.}\
  \bibnamefont {Thom}},\ }\href@noop {} {\bibfield  {journal} {\bibinfo
  {journal} {J. Chem. Phys.}\ }\textbf {\bibinfo {volume} {147}},\ \bibinfo
  {pages} {124105} (\bibinfo {year} {2017})}\BibitemShut {NoStop}%
\bibitem [{\citenamefont {Neufeld}\ and\ \citenamefont
  {Thom}(2017)}]{Neufeld2017}%
  \BibitemOpen
  \bibfield  {author} {\bibinfo {author} {\bibfnamefont {V.~A.}\ \bibnamefont
  {Neufeld}}\ and\ \bibinfo {author} {\bibfnamefont {A.~J.~W.}\ \bibnamefont
  {Thom}},\ }\href@noop {} {\bibfield  {journal} {\bibinfo  {journal} {J. Chem.
  Phys.}\ }\textbf {\bibinfo {volume} {147}},\ \bibinfo {pages} {194105}
  (\bibinfo {year} {2017})}\BibitemShut {NoStop}%
\bibitem [{\citenamefont {Blunt}\ \emph {et~al.}(2014)\citenamefont {Blunt},
  \citenamefont {Rogers}, \citenamefont {Spencer},\ and\ \citenamefont
  {Foulkes}}]{Blunt2014}%
  \BibitemOpen
  \bibfield  {author} {\bibinfo {author} {\bibfnamefont {N.~S.}\ \bibnamefont
  {Blunt}}, \bibinfo {author} {\bibfnamefont {T.~W.}\ \bibnamefont {Rogers}},
  \bibinfo {author} {\bibfnamefont {J.~S.}\ \bibnamefont {Spencer}}, \ and\
  \bibinfo {author} {\bibfnamefont {W.~M.~C.}\ \bibnamefont {Foulkes}},\
  }\href@noop {} {\bibfield  {journal} {\bibinfo  {journal} {Phys. Rev. B}\
  }\textbf {\bibinfo {volume} {89}},\ \bibinfo {pages} {245124} (\bibinfo
  {year} {2014})}\BibitemShut {NoStop}%
\bibitem [{\citenamefont {Malone}\ \emph {et~al.}(2015)\citenamefont {Malone},
  \citenamefont {Blunt}, \citenamefont {Shepherd}, \citenamefont {Lee},
  \citenamefont {Spencer},\ and\ \citenamefont {Foulkes}}]{Malone2015}%
  \BibitemOpen
  \bibfield  {author} {\bibinfo {author} {\bibfnamefont {F.~D.}\ \bibnamefont
  {Malone}}, \bibinfo {author} {\bibfnamefont {N.~S.}\ \bibnamefont {Blunt}},
  \bibinfo {author} {\bibfnamefont {J.~J.}\ \bibnamefont {Shepherd}}, \bibinfo
  {author} {\bibfnamefont {D.~K.~K.}\ \bibnamefont {Lee}}, \bibinfo {author}
  {\bibfnamefont {J.~S.}\ \bibnamefont {Spencer}}, \ and\ \bibinfo {author}
  {\bibfnamefont {W.~M.~C.}\ \bibnamefont {Foulkes}},\ }\href@noop {}
  {\bibfield  {journal} {\bibinfo  {journal} {J. Chem. Phys.}\ }\textbf
  {\bibinfo {volume} {143}},\ \bibinfo {pages} {044116} (\bibinfo {year}
  {2015})}\BibitemShut {NoStop}%
\bibitem [{\citenamefont {Malone}\ \emph {et~al.}(2016)\citenamefont {Malone},
  \citenamefont {Blunt}, \citenamefont {Brown}, \citenamefont {Lee},
  \citenamefont {Spencer}, \citenamefont {Foulkes},\ and\ \citenamefont
  {Shepherd}}]{Malone2016}%
  \BibitemOpen
  \bibfield  {author} {\bibinfo {author} {\bibfnamefont {F.~D.}\ \bibnamefont
  {Malone}}, \bibinfo {author} {\bibfnamefont {N.~S.}\ \bibnamefont {Blunt}},
  \bibinfo {author} {\bibfnamefont {E.~W.}\ \bibnamefont {Brown}}, \bibinfo
  {author} {\bibfnamefont {D.~K.~K.}\ \bibnamefont {Lee}}, \bibinfo {author}
  {\bibfnamefont {J.~S.}\ \bibnamefont {Spencer}}, \bibinfo {author}
  {\bibfnamefont {W.~M.~C.}\ \bibnamefont {Foulkes}}, \ and\ \bibinfo {author}
  {\bibfnamefont {J.~J.}\ \bibnamefont {Shepherd}},\ }\href@noop {} {\bibfield
  {journal} {\bibinfo  {journal} {Phys. Rev. Lett.}\ }\textbf {\bibinfo
  {volume} {117}},\ \bibinfo {pages} {115701} (\bibinfo {year}
  {2016})}\BibitemShut {NoStop}%
\bibitem [{\citenamefont {Ten-no}(2013)}]{Ten-no2013}%
  \BibitemOpen
  \bibfield  {author} {\bibinfo {author} {\bibfnamefont {S.}~\bibnamefont
  {Ten-no}},\ }\href@noop {} {\bibfield  {journal} {\bibinfo  {journal} {J.
  Chem. Phys.}\ }\textbf {\bibinfo {volume} {138}},\ \bibinfo {pages} {164126}
  (\bibinfo {year} {2013})}\BibitemShut {NoStop}%
\bibitem [{\citenamefont {Ohtsuka}\ and\ \citenamefont
  {Ten-no}(2015)}]{Ohtsuka2015}%
  \BibitemOpen
  \bibfield  {author} {\bibinfo {author} {\bibfnamefont {Y.}~\bibnamefont
  {Ohtsuka}}\ and\ \bibinfo {author} {\bibfnamefont {S.}~\bibnamefont
  {Ten-no}},\ }\href@noop {} {\bibfield  {journal} {\bibinfo  {journal} {J.
  Chem. Phys.}\ }\textbf {\bibinfo {volume} {143}},\ \bibinfo {pages} {214107}
  (\bibinfo {year} {2015})}\BibitemShut {NoStop}%
\bibitem [{\citenamefont {Ten-no}(2017)}]{Ten-no2017}%
  \BibitemOpen
  \bibfield  {author} {\bibinfo {author} {\bibfnamefont {S.}~\bibnamefont
  {Ten-no}},\ }\href@noop {} {\bibfield  {journal} {\bibinfo  {journal} {J.
  Chem. Phys.}\ }\textbf {\bibinfo {volume} {147}},\ \bibinfo {pages} {244107}
  (\bibinfo {year} {2017})}\BibitemShut {NoStop}%
\bibitem [{\citenamefont {Petruzielo}\ \emph {et~al.}(2012)\citenamefont
  {Petruzielo}, \citenamefont {Holmes}, \citenamefont {Changlani},
  \citenamefont {Nightingale},\ and\ \citenamefont {Umrigar}}]{Petruzielo2012}%
  \BibitemOpen
  \bibfield  {author} {\bibinfo {author} {\bibfnamefont {F.~R.}\ \bibnamefont
  {Petruzielo}}, \bibinfo {author} {\bibfnamefont {A.~A.}\ \bibnamefont
  {Holmes}}, \bibinfo {author} {\bibfnamefont {H.~J.}\ \bibnamefont
  {Changlani}}, \bibinfo {author} {\bibfnamefont {M.~P.}\ \bibnamefont
  {Nightingale}}, \ and\ \bibinfo {author} {\bibfnamefont {C.~J.}\ \bibnamefont
  {Umrigar}},\ }\href@noop {} {\bibfield  {journal} {\bibinfo  {journal} {Phys.
  Rev. Lett.}\ }\textbf {\bibinfo {volume} {109}},\ \bibinfo {pages} {230201}
  (\bibinfo {year} {2012})}\BibitemShut {NoStop}%
\bibitem [{\citenamefont {Blunt}\ \emph
  {et~al.}(2015{\natexlab{a}})\citenamefont {Blunt}, \citenamefont {Smart},
  \citenamefont {Kersten}, \citenamefont {Spencer}, \citenamefont {Booth},\
  and\ \citenamefont {Alavi}}]{Blunt2015}%
  \BibitemOpen
  \bibfield  {author} {\bibinfo {author} {\bibfnamefont {N.~S.}\ \bibnamefont
  {Blunt}}, \bibinfo {author} {\bibfnamefont {S.~D.}\ \bibnamefont {Smart}},
  \bibinfo {author} {\bibfnamefont {J.~A.~F.}\ \bibnamefont {Kersten}},
  \bibinfo {author} {\bibfnamefont {J.~S.}\ \bibnamefont {Spencer}}, \bibinfo
  {author} {\bibfnamefont {G.~H.}\ \bibnamefont {Booth}}, \ and\ \bibinfo
  {author} {\bibfnamefont {A.}~\bibnamefont {Alavi}},\ }\href@noop {}
  {\bibfield  {journal} {\bibinfo  {journal} {J. Chem. Phys.}\ }\textbf
  {\bibinfo {volume} {142}},\ \bibinfo {pages} {184107} (\bibinfo {year}
  {2015}{\natexlab{a}})}\BibitemShut {NoStop}%
\bibitem [{\citenamefont {Holmes}, \citenamefont {Changlani},\ and\
  \citenamefont {Umrigar}(2016)}]{Holmes2016_1}%
  \BibitemOpen
  \bibfield  {author} {\bibinfo {author} {\bibfnamefont {A.~A.}\ \bibnamefont
  {Holmes}}, \bibinfo {author} {\bibfnamefont {H.~J.}\ \bibnamefont
  {Changlani}}, \ and\ \bibinfo {author} {\bibfnamefont {C.~J.}\ \bibnamefont
  {Umrigar}},\ }\href@noop {} {\bibfield  {journal} {\bibinfo  {journal} {J.
  Chem. Theory Comput.}\ }\textbf {\bibinfo {volume} {12}},\ \bibinfo {pages}
  {1561} (\bibinfo {year} {2016})}\BibitemShut {NoStop}%
\bibitem [{\citenamefont {Cleland}, \citenamefont {Booth},\ and\ \citenamefont
  {Alavi}(2010)}]{Cleland2010}%
  \BibitemOpen
  \bibfield  {author} {\bibinfo {author} {\bibfnamefont {D.~M.}\ \bibnamefont
  {Cleland}}, \bibinfo {author} {\bibfnamefont {G.~H.}\ \bibnamefont {Booth}},
  \ and\ \bibinfo {author} {\bibfnamefont {A.}~\bibnamefont {Alavi}},\
  }\href@noop {} {\bibfield  {journal} {\bibinfo  {journal} {J. Chem. Phys.}\
  }\textbf {\bibinfo {volume} {132}},\ \bibinfo {pages} {041103} (\bibinfo
  {year} {2010})}\BibitemShut {NoStop}%
\bibitem [{\citenamefont {Cleland}, \citenamefont {Booth},\ and\ \citenamefont
  {Alavi}(2011)}]{Cleland2011}%
  \BibitemOpen
  \bibfield  {author} {\bibinfo {author} {\bibfnamefont {D.~M.}\ \bibnamefont
  {Cleland}}, \bibinfo {author} {\bibfnamefont {G.~H.}\ \bibnamefont {Booth}},
  \ and\ \bibinfo {author} {\bibfnamefont {A.}~\bibnamefont {Alavi}},\
  }\href@noop {} {\bibfield  {journal} {\bibinfo  {journal} {J. Chem. Phys.}\
  }\textbf {\bibinfo {volume} {134}},\ \bibinfo {pages} {024112} (\bibinfo
  {year} {2011})}\BibitemShut {NoStop}%
\bibitem [{\citenamefont {Spencer}, \citenamefont {Blunt},\ and\ \citenamefont
  {Foulkes}(2012)}]{Spencer2012}%
  \BibitemOpen
  \bibfield  {author} {\bibinfo {author} {\bibfnamefont {J.~S.}\ \bibnamefont
  {Spencer}}, \bibinfo {author} {\bibfnamefont {N.~S.}\ \bibnamefont {Blunt}},
  \ and\ \bibinfo {author} {\bibfnamefont {W.~M.~C.}\ \bibnamefont {Foulkes}},\
  }\href@noop {} {\bibfield  {journal} {\bibinfo  {journal} {J. Chem. Phys.}\
  }\textbf {\bibinfo {volume} {136}},\ \bibinfo {pages} {054110} (\bibinfo
  {year} {2012})}\BibitemShut {NoStop}%
\bibitem [{\citenamefont {Huron}, \citenamefont {Malrieu},\ and\ \citenamefont
  {Rancurel}(1973)}]{Huron1973}%
  \BibitemOpen
  \bibfield  {author} {\bibinfo {author} {\bibfnamefont {B.}~\bibnamefont
  {Huron}}, \bibinfo {author} {\bibfnamefont {J.~P.}\ \bibnamefont {Malrieu}},
  \ and\ \bibinfo {author} {\bibfnamefont {P.}~\bibnamefont {Rancurel}},\
  }\href@noop {} {\bibfield  {journal} {\bibinfo  {journal} {J. Chem. Phys.}\
  }\textbf {\bibinfo {volume} {58}},\ \bibinfo {pages} {5745} (\bibinfo {year}
  {1973})}\BibitemShut {NoStop}%
\bibitem [{\citenamefont {Buenker}\ and\ \citenamefont
  {Peyerimhoff}(1974)}]{Buenker1974}%
  \BibitemOpen
  \bibfield  {author} {\bibinfo {author} {\bibfnamefont {R.~J.}\ \bibnamefont
  {Buenker}}\ and\ \bibinfo {author} {\bibfnamefont {S.~D.}\ \bibnamefont
  {Peyerimhoff}},\ }\href@noop {} {\bibfield  {journal} {\bibinfo  {journal}
  {Theor. Chim. Acta}\ }\textbf {\bibinfo {volume} {35}},\ \bibinfo {pages}
  {33} (\bibinfo {year} {1974})}\BibitemShut {NoStop}%
\bibitem [{\citenamefont {Liu}\ and\ \citenamefont {Hoffman}(2016)}]{Liu2016}%
  \BibitemOpen
  \bibfield  {author} {\bibinfo {author} {\bibfnamefont {W.}~\bibnamefont
  {Liu}}\ and\ \bibinfo {author} {\bibfnamefont {M.~R.}\ \bibnamefont
  {Hoffman}},\ }\href@noop {} {\bibfield  {journal} {\bibinfo  {journal} {J.
  Chem. Theory Comput.}\ }\textbf {\bibinfo {volume} {12}},\ \bibinfo {pages}
  {1169} (\bibinfo {year} {2016})}\BibitemShut {NoStop}%
\bibitem [{\citenamefont {Schriber}\ and\ \citenamefont
  {Evangelista}(2016)}]{Schriber2016}%
  \BibitemOpen
  \bibfield  {author} {\bibinfo {author} {\bibfnamefont {J.~B.}\ \bibnamefont
  {Schriber}}\ and\ \bibinfo {author} {\bibfnamefont {F.~A.}\ \bibnamefont
  {Evangelista}},\ }\href@noop {} {\bibfield  {journal} {\bibinfo  {journal}
  {J. Chem. Phys.}\ }\textbf {\bibinfo {volume} {144}},\ \bibinfo {pages}
  {161106} (\bibinfo {year} {2016})}\BibitemShut {NoStop}%
\bibitem [{\citenamefont {Tubman}\ \emph {et~al.}(2016)\citenamefont {Tubman},
  \citenamefont {Lee}, \citenamefont {Takeshita}, \citenamefont {Head-Gordon},\
  and\ \citenamefont {Whaley}}]{Tubman2016}%
  \BibitemOpen
  \bibfield  {author} {\bibinfo {author} {\bibfnamefont {N.~M.}\ \bibnamefont
  {Tubman}}, \bibinfo {author} {\bibfnamefont {J.}~\bibnamefont {Lee}},
  \bibinfo {author} {\bibfnamefont {T.~Y.}\ \bibnamefont {Takeshita}}, \bibinfo
  {author} {\bibfnamefont {M.}~\bibnamefont {Head-Gordon}}, \ and\ \bibinfo
  {author} {\bibfnamefont {B.}~\bibnamefont {Whaley}},\ }\href@noop {}
  {\bibfield  {journal} {\bibinfo  {journal} {J. Chem. Phys.}\ }\textbf
  {\bibinfo {volume} {145}},\ \bibinfo {pages} {044112} (\bibinfo {year}
  {2016})}\BibitemShut {NoStop}%
\bibitem [{\citenamefont {Holmes}, \citenamefont {Tubman},\ and\ \citenamefont
  {Umrigar}(2016)}]{Holmes2016_2}%
  \BibitemOpen
  \bibfield  {author} {\bibinfo {author} {\bibfnamefont {A.~A.}\ \bibnamefont
  {Holmes}}, \bibinfo {author} {\bibfnamefont {N.~M.}\ \bibnamefont {Tubman}},
  \ and\ \bibinfo {author} {\bibfnamefont {C.~J.}\ \bibnamefont {Umrigar}},\
  }\href@noop {} {\bibfield  {journal} {\bibinfo  {journal} {J. Chem. Theory
  Comput.}\ }\textbf {\bibinfo {volume} {12}},\ \bibinfo {pages} {3674}
  (\bibinfo {year} {2016})}\BibitemShut {NoStop}%
\bibitem [{\citenamefont {Sharma}\ \emph {et~al.}(2017)\citenamefont {Sharma},
  \citenamefont {Holmes}, \citenamefont {Jeanmairet}, \citenamefont {Alavi},\
  and\ \citenamefont {Umrigar}}]{Sharma2017}%
  \BibitemOpen
  \bibfield  {author} {\bibinfo {author} {\bibfnamefont {S.}~\bibnamefont
  {Sharma}}, \bibinfo {author} {\bibfnamefont {A.~A.}\ \bibnamefont {Holmes}},
  \bibinfo {author} {\bibfnamefont {G.}~\bibnamefont {Jeanmairet}}, \bibinfo
  {author} {\bibfnamefont {A.}~\bibnamefont {Alavi}}, \ and\ \bibinfo {author}
  {\bibfnamefont {C.~J.}\ \bibnamefont {Umrigar}},\ }\href@noop {} {\bibfield
  {journal} {\bibinfo  {journal} {J. Chem. Theory Comput.}\ }\textbf {\bibinfo
  {volume} {13}},\ \bibinfo {pages} {1595} (\bibinfo {year}
  {2017})}\BibitemShut {NoStop}%
\bibitem [{\citenamefont {Garniron}\ \emph {et~al.}(2017)\citenamefont
  {Garniron}, \citenamefont {Scemama}, \citenamefont {Loos},\ and\
  \citenamefont {Caffarel}}]{Garniron2017}%
  \BibitemOpen
  \bibfield  {author} {\bibinfo {author} {\bibfnamefont {Y.}~\bibnamefont
  {Garniron}}, \bibinfo {author} {\bibfnamefont {A.}~\bibnamefont {Scemama}},
  \bibinfo {author} {\bibfnamefont {P.-F.}\ \bibnamefont {Loos}}, \ and\
  \bibinfo {author} {\bibfnamefont {M.}~\bibnamefont {Caffarel}},\ }\href@noop
  {} {\bibfield  {journal} {\bibinfo  {journal} {J. Chem. Phys.}\ }\textbf
  {\bibinfo {volume} {147}},\ \bibinfo {pages} {034101} (\bibinfo {year}
  {2017})}\BibitemShut {NoStop}%
\bibitem [{\citenamefont {Evangelisti}, \citenamefont {Daudey},\ and\
  \citenamefont {Malrieu}(1983)}]{Evangelisti1983}%
  \BibitemOpen
  \bibfield  {author} {\bibinfo {author} {\bibfnamefont {S.}~\bibnamefont
  {Evangelisti}}, \bibinfo {author} {\bibfnamefont {J.-P.}\ \bibnamefont
  {Daudey}}, \ and\ \bibinfo {author} {\bibfnamefont {J.-P.}\ \bibnamefont
  {Malrieu}},\ }\href@noop {} {\bibfield  {journal} {\bibinfo  {journal} {Chem.
  Phys.}\ }\textbf {\bibinfo {volume} {75}},\ \bibinfo {pages} {91} (\bibinfo
  {year} {1983})}\BibitemShut {NoStop}%
\bibitem [{\citenamefont {Giner}, \citenamefont {Scemama},\ and\ \citenamefont
  {Caffarel}(2013)}]{Scemama2013}%
  \BibitemOpen
  \bibfield  {author} {\bibinfo {author} {\bibfnamefont {E.}~\bibnamefont
  {Giner}}, \bibinfo {author} {\bibfnamefont {A.}~\bibnamefont {Scemama}}, \
  and\ \bibinfo {author} {\bibfnamefont {M.}~\bibnamefont {Caffarel}},\
  }\href@noop {} {\bibfield  {journal} {\bibinfo  {journal} {Can. J. Chem.}\
  }\textbf {\bibinfo {volume} {91}},\ \bibinfo {pages} {879} (\bibinfo {year}
  {2013})}\BibitemShut {NoStop}%
\bibitem [{\citenamefont {Scemama}\ \emph {et~al.}(2014)\citenamefont
  {Scemama}, \citenamefont {Applencourt}, \citenamefont {Giner},\ and\
  \citenamefont {Caffarel}}]{Scemama2014}%
  \BibitemOpen
  \bibfield  {author} {\bibinfo {author} {\bibfnamefont {A.}~\bibnamefont
  {Scemama}}, \bibinfo {author} {\bibfnamefont {T.}~\bibnamefont
  {Applencourt}}, \bibinfo {author} {\bibfnamefont {E.}~\bibnamefont {Giner}},
  \ and\ \bibinfo {author} {\bibfnamefont {M.}~\bibnamefont {Caffarel}},\
  }\href@noop {} {\bibfield  {journal} {\bibinfo  {journal} {J. Chem. Phys.}\
  }\textbf {\bibinfo {volume} {141}},\ \bibinfo {pages} {244110} (\bibinfo
  {year} {2014})}\BibitemShut {NoStop}%
\bibitem [{\citenamefont {Caffarel}\ \emph {et~al.}(2016)\citenamefont
  {Caffarel}, \citenamefont {Applencourt}, \citenamefont {Giner},\ and\
  \citenamefont {Scemama}}]{Caffarel2016}%
  \BibitemOpen
  \bibfield  {author} {\bibinfo {author} {\bibfnamefont {M.}~\bibnamefont
  {Caffarel}}, \bibinfo {author} {\bibfnamefont {T.}~\bibnamefont
  {Applencourt}}, \bibinfo {author} {\bibfnamefont {E.}~\bibnamefont {Giner}},
  \ and\ \bibinfo {author} {\bibfnamefont {A.}~\bibnamefont {Scemama}},\ }in\
  \href@noop {} {\emph {\bibinfo {booktitle} {Recent Progress in Quantum Monte
  Carlo}}}\ (\bibinfo  {publisher} {American Physical Society},\ \bibinfo
  {year} {2016})\ Chap.~\bibinfo {chapter} {2}, pp.\ \bibinfo {pages}
  {15--46}\BibitemShut {NoStop}%
\bibitem [{\citenamefont {Smith}\ \emph {et~al.}(2017)\citenamefont {Smith},
  \citenamefont {Mussard}, \citenamefont {Holmes},\ and\ \citenamefont
  {Sharma}}]{Smith2017}%
  \BibitemOpen
  \bibfield  {author} {\bibinfo {author} {\bibfnamefont {J.~E.~T.}\
  \bibnamefont {Smith}}, \bibinfo {author} {\bibfnamefont {B.}~\bibnamefont
  {Mussard}}, \bibinfo {author} {\bibfnamefont {A.~A.}\ \bibnamefont {Holmes}},
  \ and\ \bibinfo {author} {\bibfnamefont {S.}~\bibnamefont {Sharma}},\
  }\href@noop {} {\bibfield  {journal} {\bibinfo  {journal} {J. Chem. Theory
  Comput.}\ }\textbf {\bibinfo {volume} {13}},\ \bibinfo {pages} {5468}
  (\bibinfo {year} {2017})}\BibitemShut {NoStop}%
\bibitem [{\citenamefont {Holmes}, \citenamefont {Umrigar},\ and\ \citenamefont
  {Sharma}(2017)}]{Holmes2017}%
  \BibitemOpen
  \bibfield  {author} {\bibinfo {author} {\bibfnamefont {A.~A.}\ \bibnamefont
  {Holmes}}, \bibinfo {author} {\bibfnamefont {C.~J.}\ \bibnamefont {Umrigar}},
  \ and\ \bibinfo {author} {\bibfnamefont {S.}~\bibnamefont {Sharma}},\
  }\href@noop {} {\bibfield  {journal} {\bibinfo  {journal} {J. Chem. Phys.}\
  }\textbf {\bibinfo {volume} {147}},\ \bibinfo {pages} {164111} (\bibinfo
  {year} {2017})}\BibitemShut {NoStop}%
\bibitem [{\citenamefont {Chien}\ \emph {et~al.}(2018)\citenamefont {Chien},
  \citenamefont {Holmes}, \citenamefont {Otten}, \citenamefont {Umrigar},
  \citenamefont {Sharma},\ and\ \citenamefont {Zimmerman}}]{Chien2018}%
  \BibitemOpen
  \bibfield  {author} {\bibinfo {author} {\bibfnamefont {A.~D.}\ \bibnamefont
  {Chien}}, \bibinfo {author} {\bibfnamefont {A.~A.}\ \bibnamefont {Holmes}},
  \bibinfo {author} {\bibfnamefont {M.}~\bibnamefont {Otten}}, \bibinfo
  {author} {\bibfnamefont {C.~J.}\ \bibnamefont {Umrigar}}, \bibinfo {author}
  {\bibfnamefont {S.}~\bibnamefont {Sharma}}, \ and\ \bibinfo {author}
  {\bibfnamefont {P.~M.}\ \bibnamefont {Zimmerman}},\ }\href@noop {} {\bibfield
   {journal} {\bibinfo  {journal} {J. Phys. Chem. A}\ }\textbf {\bibinfo
  {volume} {122}},\ \bibinfo {pages} {2714} (\bibinfo {year}
  {2018})}\BibitemShut {NoStop}%
\bibitem [{\citenamefont {Scemama}\ \emph {et~al.}(2018)\citenamefont
  {Scemama}, \citenamefont {Garniron}, \citenamefont {Caffarel},\ and\
  \citenamefont {Loos}}]{Scemama2018}%
  \BibitemOpen
  \bibfield  {author} {\bibinfo {author} {\bibfnamefont {A.}~\bibnamefont
  {Scemama}}, \bibinfo {author} {\bibfnamefont {Y.}~\bibnamefont {Garniron}},
  \bibinfo {author} {\bibfnamefont {M.}~\bibnamefont {Caffarel}}, \ and\
  \bibinfo {author} {\bibfnamefont {P.-F.}\ \bibnamefont {Loos}},\ }\href@noop
  {} {\bibfield  {journal} {\bibinfo  {journal} {J. Chem. Theory Comput.}\
  }\textbf {\bibinfo {volume} {14}},\ \bibinfo {pages} {1395} (\bibinfo {year}
  {2018})}\BibitemShut {NoStop}%
\bibitem [{\citenamefont {Dash}\ \emph {et~al.}(2018)\citenamefont {Dash},
  \citenamefont {Moroni}, \citenamefont {Scemama},\ and\ \citenamefont
  {Filippi}}]{Dash2018}%
  \BibitemOpen
  \bibfield  {author} {\bibinfo {author} {\bibfnamefont {M.}~\bibnamefont
  {Dash}}, \bibinfo {author} {\bibfnamefont {S.}~\bibnamefont {Moroni}},
  \bibinfo {author} {\bibfnamefont {A.}~\bibnamefont {Scemama}}, \ and\
  \bibinfo {author} {\bibfnamefont {C.}~\bibnamefont {Filippi}},\ }\href@noop
  {} {\bibfield  {journal} {\bibinfo  {journal} {arXiv:1804.09610
  [physics.chem-ph]}\ } (\bibinfo {year} {2018})}\BibitemShut {NoStop}%
\bibitem [{\citenamefont {Sharma}(2018)}]{Sharma2018}%
  \BibitemOpen
  \bibfield  {author} {\bibinfo {author} {\bibfnamefont {S.}~\bibnamefont
  {Sharma}},\ }\href@noop {} {\bibfield  {journal} {\bibinfo  {journal}
  {arXiv:1803.04341 [cond-mat.str-el]}\ } (\bibinfo {year} {2018})}\BibitemShut
  {NoStop}%
\bibitem [{\citenamefont {Guo}, \citenamefont {Li},\ and\ \citenamefont
  {Chan}(2018{\natexlab{a}})}]{Guo2018_1}%
  \BibitemOpen
  \bibfield  {author} {\bibinfo {author} {\bibfnamefont {S.}~\bibnamefont
  {Guo}}, \bibinfo {author} {\bibfnamefont {Z.}~\bibnamefont {Li}}, \ and\
  \bibinfo {author} {\bibfnamefont {G.~K.-L.}\ \bibnamefont {Chan}},\
  }\href@noop {} {\bibfield  {journal} {\bibinfo  {journal} {arXiv:1803.07150
  [physics.chem-ph]}\ } (\bibinfo {year} {2018}{\natexlab{a}})}\BibitemShut
  {NoStop}%
\bibitem [{\citenamefont {Guo}, \citenamefont {Li},\ and\ \citenamefont
  {Chan}(2018{\natexlab{b}})}]{Guo2018_2}%
  \BibitemOpen
  \bibfield  {author} {\bibinfo {author} {\bibfnamefont {S.}~\bibnamefont
  {Guo}}, \bibinfo {author} {\bibfnamefont {Z.}~\bibnamefont {Li}}, \ and\
  \bibinfo {author} {\bibfnamefont {G.~K.-L.}\ \bibnamefont {Chan}},\
  }\href@noop {} {\bibfield  {journal} {\bibinfo  {journal} {arXiv:1803.09943
  [physics.chem-ph]}\ } (\bibinfo {year} {2018}{\natexlab{b}})}\BibitemShut
  {NoStop}%
\bibitem [{fn2()}]{fn2}%
  \BibitemOpen
  \href@noop {} {}\bibinfo {note} {For an FCIQMC wave function this expectation
  value is defined as $\textrm{E}[ \;| \Psi(\tau) \ket \;] = \sum_{\alpha}
  p_{\alpha} | \tilde{\Psi}(\alpha; \tau) \ket$, where $ | \tilde{\Psi}(\alpha;
  \tau) \ket $ denotes a possible wave function at $\tau$, and $p_{\alpha}$
  denotes the probability of it having been selected.}\BibitemShut {Stop}%
\bibitem [{fn1()}]{fn1}%
  \BibitemOpen
  \href@noop {} {}\bibinfo {note} {Early descriptions of i-FCIQMC state that
  the two spawnings must also have the same sign, although this is no longer
  applied in \url{NECI} - spawnings of opposite signs are allowed to survive
  also, with only very slight changes to results.}\BibitemShut {Stop}%
\bibitem [{\citenamefont {Zhang}\ and\ \citenamefont
  {Kalos}(1993)}]{Zhang1993}%
  \BibitemOpen
  \bibfield  {author} {\bibinfo {author} {\bibfnamefont {S.}~\bibnamefont
  {Zhang}}\ and\ \bibinfo {author} {\bibfnamefont {M.~H.}\ \bibnamefont
  {Kalos}},\ }\href@noop {} {\bibfield  {journal} {\bibinfo  {journal} {J.
  Stat. Phys.}\ }\textbf {\bibinfo {volume} {70}},\ \bibinfo {pages} {515}
  (\bibinfo {year} {1993})}\BibitemShut {NoStop}%
\bibitem [{\citenamefont {Hastings}\ \emph {et~al.}(2010)\citenamefont
  {Hastings}, \citenamefont {Gonz\'{a}lez}, \citenamefont {Kallin},\ and\
  \citenamefont {Melko}}]{Hastings2010}%
  \BibitemOpen
  \bibfield  {author} {\bibinfo {author} {\bibfnamefont {M.~B.}\ \bibnamefont
  {Hastings}}, \bibinfo {author} {\bibfnamefont {I.}~\bibnamefont
  {Gonz\'{a}lez}}, \bibinfo {author} {\bibfnamefont {A.~B.}\ \bibnamefont
  {Kallin}}, \ and\ \bibinfo {author} {\bibfnamefont {R.~G.}\ \bibnamefont
  {Melko}},\ }\href@noop {} {\bibfield  {journal} {\bibinfo  {journal} {Phys.
  Rev. Lett.}\ }\textbf {\bibinfo {volume} {104}},\ \bibinfo {pages} {157201}
  (\bibinfo {year} {2010})}\BibitemShut {NoStop}%
\bibitem [{\citenamefont {Overy}\ \emph {et~al.}(2014)\citenamefont {Overy},
  \citenamefont {Booth}, \citenamefont {Blunt}, \citenamefont {Shepherd},
  \citenamefont {Cleland},\ and\ \citenamefont {Alavi}}]{Overy2014}%
  \BibitemOpen
  \bibfield  {author} {\bibinfo {author} {\bibfnamefont {C.}~\bibnamefont
  {Overy}}, \bibinfo {author} {\bibfnamefont {G.~H.}\ \bibnamefont {Booth}},
  \bibinfo {author} {\bibfnamefont {N.~S.}\ \bibnamefont {Blunt}}, \bibinfo
  {author} {\bibfnamefont {J.~J.}\ \bibnamefont {Shepherd}}, \bibinfo {author}
  {\bibfnamefont {D.}~\bibnamefont {Cleland}}, \ and\ \bibinfo {author}
  {\bibfnamefont {A.}~\bibnamefont {Alavi}},\ }\href@noop {} {\bibfield
  {journal} {\bibinfo  {journal} {J. Chem. Phys.}\ }\textbf {\bibinfo {volume}
  {141}},\ \bibinfo {pages} {244117} (\bibinfo {year} {2014})}\BibitemShut
  {NoStop}%
\bibitem [{\citenamefont {Blunt}\ \emph
  {et~al.}(2015{\natexlab{b}})\citenamefont {Blunt}, \citenamefont {Smart},
  \citenamefont {Booth},\ and\ \citenamefont {Alavi}}]{Blunt2015_3}%
  \BibitemOpen
  \bibfield  {author} {\bibinfo {author} {\bibfnamefont {N.~S.}\ \bibnamefont
  {Blunt}}, \bibinfo {author} {\bibfnamefont {S.~D.}\ \bibnamefont {Smart}},
  \bibinfo {author} {\bibfnamefont {G.~H.}\ \bibnamefont {Booth}}, \ and\
  \bibinfo {author} {\bibfnamefont {A.}~\bibnamefont {Alavi}},\ }\href@noop {}
  {\bibfield  {journal} {\bibinfo  {journal} {J. Chem. Phys.}\ }\textbf
  {\bibinfo {volume} {143}},\ \bibinfo {pages} {134117} (\bibinfo {year}
  {2015}{\natexlab{b}})}\BibitemShut {NoStop}%
\bibitem [{\citenamefont {Blunt}, \citenamefont {Booth},\ and\ \citenamefont
  {Alavi}(2017)}]{Blunt2017}%
  \BibitemOpen
  \bibfield  {author} {\bibinfo {author} {\bibfnamefont {N.~S.}\ \bibnamefont
  {Blunt}}, \bibinfo {author} {\bibfnamefont {G.~H.}\ \bibnamefont {Booth}}, \
  and\ \bibinfo {author} {\bibfnamefont {A.}~\bibnamefont {Alavi}},\
  }\href@noop {} {\bibfield  {journal} {\bibinfo  {journal} {J. Chem. Phys.}\
  }\textbf {\bibinfo {volume} {146}},\ \bibinfo {pages} {244105} (\bibinfo
  {year} {2017})}\BibitemShut {NoStop}%
\bibitem [{NEC()}]{NECI_github}%
  \BibitemOpen
  \href@noop {} {\enquote {\bibinfo {title} {Neci github web page},}\ }\bibinfo
  {howpublished} {\url{https://github.com/ghb24/NECI_STABLE}}\BibitemShut
  {NoStop}%
\bibitem [{\citenamefont {Sun}\ \emph {et~al.}(2017)\citenamefont {Sun},
  \citenamefont {Berkelbach}, \citenamefont {Blunt}, \citenamefont {Booth},
  \citenamefont {Guo}, \citenamefont {Li}, \citenamefont {Liu}, \citenamefont
  {McClain}, \citenamefont {Sharma}, \citenamefont {Wouters},\ and\
  \citenamefont {Chan}}]{pyscf}%
  \BibitemOpen
  \bibfield  {author} {\bibinfo {author} {\bibfnamefont {Q.}~\bibnamefont
  {Sun}}, \bibinfo {author} {\bibfnamefont {T.~C.}\ \bibnamefont {Berkelbach}},
  \bibinfo {author} {\bibfnamefont {N.~S.}\ \bibnamefont {Blunt}}, \bibinfo
  {author} {\bibfnamefont {G.~H.}\ \bibnamefont {Booth}}, \bibinfo {author}
  {\bibfnamefont {S.}~\bibnamefont {Guo}}, \bibinfo {author} {\bibfnamefont
  {Z.}~\bibnamefont {Li}}, \bibinfo {author} {\bibfnamefont {J.}~\bibnamefont
  {Liu}}, \bibinfo {author} {\bibfnamefont {J.}~\bibnamefont {McClain}},
  \bibinfo {author} {\bibfnamefont {S.}~\bibnamefont {Sharma}}, \bibinfo
  {author} {\bibfnamefont {S.}~\bibnamefont {Wouters}}, \ and\ \bibinfo
  {author} {\bibfnamefont {G.~K.-L.}\ \bibnamefont {Chan}},\ }\href@noop {}
  {\bibfield  {journal} {\bibinfo  {journal} {WIREs Comput Mol Sci 2018}\
  }\textbf {\bibinfo {volume} {8}},\ \bibinfo {pages} {e1340} (\bibinfo {year}
  {2017})}\BibitemShut {NoStop}%
\bibitem [{\citenamefont {Smeyers}\ and\ \citenamefont
  {Doreste-Suarez}(1973)}]{Smeyers1973}%
  \BibitemOpen
  \bibfield  {author} {\bibinfo {author} {\bibfnamefont {Y.~G.}\ \bibnamefont
  {Smeyers}}\ and\ \bibinfo {author} {\bibfnamefont {L.}~\bibnamefont
  {Doreste-Suarez}},\ }\href@noop {} {\bibfield  {journal} {\bibinfo  {journal}
  {Int. J. Quantum Chem.}\ }\textbf {\bibinfo {volume} {7}},\ \bibinfo {pages}
  {687} (\bibinfo {year} {1973})}\BibitemShut {NoStop}%
\bibitem [{\citenamefont {Sharma}(2015)}]{Sharma2015}%
  \BibitemOpen
  \bibfield  {author} {\bibinfo {author} {\bibfnamefont {S.}~\bibnamefont
  {Sharma}},\ }\href@noop {} {\bibfield  {journal} {\bibinfo  {journal} {J.
  Chem. Phys.}\ }\textbf {\bibinfo {volume} {142}},\ \bibinfo {pages} {024107}
  (\bibinfo {year} {2015})}\BibitemShut {NoStop}%
\bibitem [{\citenamefont {Hoy}\ and\ \citenamefont {Bunker}(1979)}]{Hoy1979}%
  \BibitemOpen
  \bibfield  {author} {\bibinfo {author} {\bibfnamefont {A.~R.}\ \bibnamefont
  {Hoy}}\ and\ \bibinfo {author} {\bibfnamefont {P.~R.}\ \bibnamefont
  {Bunker}},\ }\href@noop {} {\bibfield  {journal} {\bibinfo  {journal} {J.
  Mol. Spectrosc.}\ }\textbf {\bibinfo {volume} {74}},\ \bibinfo {pages} {1}
  (\bibinfo {year} {1979})}\BibitemShut {NoStop}%
\bibitem [{\citenamefont {Schreiber}\ \emph {et~al.}(2008)\citenamefont
  {Schreiber}, \citenamefont {Silva-Junior}, \citenamefont {Sauer},\ and\
  \citenamefont {Thiel}}]{Schreiber2008}%
  \BibitemOpen
  \bibfield  {author} {\bibinfo {author} {\bibfnamefont {M.}~\bibnamefont
  {Schreiber}}, \bibinfo {author} {\bibfnamefont {M.~R.}\ \bibnamefont
  {Silva-Junior}}, \bibinfo {author} {\bibfnamefont {S.~P.~A.}\ \bibnamefont
  {Sauer}}, \ and\ \bibinfo {author} {\bibfnamefont {W.}~\bibnamefont
  {Thiel}},\ }\href@noop {} {\bibfield  {journal} {\bibinfo  {journal} {J.
  Chem. Phys.}\ }\textbf {\bibinfo {volume} {128}},\ \bibinfo {pages} {134110}
  (\bibinfo {year} {2008})}\BibitemShut {NoStop}%
\bibitem [{\citenamefont {Daday}\ \emph {et~al.}(2012)\citenamefont {Daday},
  \citenamefont {Smart}, \citenamefont {Booth}, \citenamefont {Alavi},\ and\
  \citenamefont {Filippi}}]{Daday2012}%
  \BibitemOpen
  \bibfield  {author} {\bibinfo {author} {\bibfnamefont {C.}~\bibnamefont
  {Daday}}, \bibinfo {author} {\bibfnamefont {S.}~\bibnamefont {Smart}},
  \bibinfo {author} {\bibfnamefont {G.~H.}\ \bibnamefont {Booth}}, \bibinfo
  {author} {\bibfnamefont {A.}~\bibnamefont {Alavi}}, \ and\ \bibinfo {author}
  {\bibfnamefont {C.}~\bibnamefont {Filippi}},\ }\href@noop {} {\bibfield
  {journal} {\bibinfo  {journal} {J. Chem. Theory Comput.}\ }\textbf {\bibinfo
  {volume} {8}},\ \bibinfo {pages} {4441} (\bibinfo {year} {2012})}\BibitemShut
  {NoStop}%
\bibitem [{\citenamefont {Olivares-Amaya}\ \emph {et~al.}(2015)\citenamefont
  {Olivares-Amaya}, \citenamefont {Hu}, \citenamefont {Nakatani}, \citenamefont
  {Sharma}, \citenamefont {Yang},\ and\ \citenamefont {Chan}}]{Olivares2015}%
  \BibitemOpen
  \bibfield  {author} {\bibinfo {author} {\bibfnamefont {R.}~\bibnamefont
  {Olivares-Amaya}}, \bibinfo {author} {\bibfnamefont {W.}~\bibnamefont {Hu}},
  \bibinfo {author} {\bibfnamefont {N.}~\bibnamefont {Nakatani}}, \bibinfo
  {author} {\bibfnamefont {S.}~\bibnamefont {Sharma}}, \bibinfo {author}
  {\bibfnamefont {J.}~\bibnamefont {Yang}}, \ and\ \bibinfo {author}
  {\bibfnamefont {G.~K.-L.}\ \bibnamefont {Chan}},\ }\href@noop {} {\bibfield
  {journal} {\bibinfo  {journal} {J. Chem. Phys.}\ }\textbf {\bibinfo {volume}
  {142}},\ \bibinfo {pages} {034102} (\bibinfo {year} {2015})}\BibitemShut
  {NoStop}%
\end{thebibliography}

\begin{thebibliography}{11}%
\makeatletter
\providecommand \@ifxundefined [1]{%
 \@ifx{#1\undefined}
}%
\providecommand \@ifnum [1]{%
 \ifnum #1\expandafter \@firstoftwo
 \else \expandafter \@secondoftwo
 \fi
}%
\providecommand \@ifx [1]{%
 \ifx #1\expandafter \@firstoftwo
 \else \expandafter \@secondoftwo
 \fi
}%
\providecommand \natexlab [1]{#1}%
\providecommand \enquote  [1]{``#1''}%
\providecommand \bibnamefont  [1]{#1}%
\providecommand \bibfnamefont [1]{#1}%
\providecommand \citenamefont [1]{#1}%
\providecommand \href@noop [0]{\@secondoftwo}%
\providecommand \href [0]{\begingroup \@sanitize@url \@href}%
\providecommand \@href[1]{\@@startlink{#1}\@@href}%
\providecommand \@@href[1]{\endgroup#1\@@endlink}%
\providecommand \@sanitize@url [0]{\catcode `\\12\catcode `\$12\catcode
  `\&12\catcode `\#12\catcode `\^12\catcode `\_12\catcode `\%12\relax}%
\providecommand \@@startlink[1]{}%
\providecommand \@@endlink[0]{}%
\providecommand \url  [0]{\begingroup\@sanitize@url \@url }%
\providecommand \@url [1]{\endgroup\@href {#1}{\urlprefix }}%
\providecommand \urlprefix  [0]{URL }%
\providecommand \Eprint [0]{\href }%
\providecommand \doibase [0]{http://dx.doi.org/}%
\providecommand \selectlanguage [0]{\@gobble}%
\providecommand \bibinfo  [0]{\@secondoftwo}%
\providecommand \bibfield  [0]{\@secondoftwo}%
\providecommand \translation [1]{[#1]}%
\providecommand \BibitemOpen [0]{}%
\providecommand \bibitemStop [0]{}%
\providecommand \bibitemNoStop [0]{.\EOS\space}%
\providecommand \EOS [0]{\spacefactor3000\relax}%
\providecommand \BibitemShut  [1]{\csname bibitem#1\endcsname}%
\let\auto@bib@innerbib\@empty
\bibitem [{\citenamefont {Petruzielo}\ \emph {et~al.}(2012)\citenamefont
  {Petruzielo}, \citenamefont {Holmes}, \citenamefont {Changlani},
  \citenamefont {Nightingale},\ and\ \citenamefont
  {Umrigar}}]{S_Petruzielo2012}%
  \BibitemOpen
  \bibfield  {author} {\bibinfo {author} {\bibfnamefont {F.~R.}\ \bibnamefont
  {Petruzielo}}, \bibinfo {author} {\bibfnamefont {A.~A.}\ \bibnamefont
  {Holmes}}, \bibinfo {author} {\bibfnamefont {H.~J.}\ \bibnamefont
  {Changlani}}, \bibinfo {author} {\bibfnamefont {M.~P.}\ \bibnamefont
  {Nightingale}}, \ and\ \bibinfo {author} {\bibfnamefont {C.~J.}\ \bibnamefont
  {Umrigar}},\ }\href@noop {} {\bibfield  {journal} {\bibinfo  {journal} {Phys.
  Rev. Lett.}\ }\textbf {\bibinfo {volume} {109}},\ \bibinfo {pages} {230201}
  (\bibinfo {year} {2012})}\BibitemShut {NoStop}%
\bibitem [{\citenamefont {Blunt}\ \emph
  {et~al.}(2015{\natexlab{a}})\citenamefont {Blunt}, \citenamefont {Smart},
  \citenamefont {Kersten}, \citenamefont {Spencer}, \citenamefont {Booth},\
  and\ \citenamefont {Alavi}}]{S_Blunt2015}%
  \BibitemOpen
  \bibfield  {author} {\bibinfo {author} {\bibfnamefont {N.~S.}\ \bibnamefont
  {Blunt}}, \bibinfo {author} {\bibfnamefont {S.~D.}\ \bibnamefont {Smart}},
  \bibinfo {author} {\bibfnamefont {J.~A.~F.}\ \bibnamefont {Kersten}},
  \bibinfo {author} {\bibfnamefont {J.~S.}\ \bibnamefont {Spencer}}, \bibinfo
  {author} {\bibfnamefont {G.~H.}\ \bibnamefont {Booth}}, \ and\ \bibinfo
  {author} {\bibfnamefont {A.}~\bibnamefont {Alavi}},\ }\href@noop {}
  {\bibfield  {journal} {\bibinfo  {journal} {J. Chem. Phys.}\ }\textbf
  {\bibinfo {volume} {142}},\ \bibinfo {pages} {184107} (\bibinfo {year}
  {2015}{\natexlab{a}})}\BibitemShut {NoStop}%
\bibitem [{\citenamefont {Smeyers}\ and\ \citenamefont
  {Doreste-Suarez}(1973)}]{S_Smeyers1973}%
  \BibitemOpen
  \bibfield  {author} {\bibinfo {author} {\bibfnamefont {Y.~G.}\ \bibnamefont
  {Smeyers}}\ and\ \bibinfo {author} {\bibfnamefont {L.}~\bibnamefont
  {Doreste-Suarez}},\ }\href@noop {} {\bibfield  {journal} {\bibinfo  {journal}
  {Int. J. Quantum Chem.}\ }\textbf {\bibinfo {volume} {7}},\ \bibinfo {pages}
  {687} (\bibinfo {year} {1973})}\BibitemShut {NoStop}%
\bibitem [{\citenamefont {Blunt}\ \emph
  {et~al.}(2015{\natexlab{b}})\citenamefont {Blunt}, \citenamefont {Smart},
  \citenamefont {Booth},\ and\ \citenamefont {Alavi}}]{S_Blunt2015_3}%
  \BibitemOpen
  \bibfield  {author} {\bibinfo {author} {\bibfnamefont {N.~S.}\ \bibnamefont
  {Blunt}}, \bibinfo {author} {\bibfnamefont {S.~D.}\ \bibnamefont {Smart}},
  \bibinfo {author} {\bibfnamefont {G.~H.}\ \bibnamefont {Booth}}, \ and\
  \bibinfo {author} {\bibfnamefont {A.}~\bibnamefont {Alavi}},\ }\href@noop {}
  {\bibfield  {journal} {\bibinfo  {journal} {J. Chem. Phys.}\ }\textbf
  {\bibinfo {volume} {143}},\ \bibinfo {pages} {134117} (\bibinfo {year}
  {2015}{\natexlab{b}})}\BibitemShut {NoStop}%
\bibitem [{\citenamefont {Holmes}\ \emph {et~al.}(2016)\citenamefont {Holmes},
  \citenamefont {Tubman},\ and\ \citenamefont {Umrigar}}]{S_Holmes2016_2}%
  \BibitemOpen
  \bibfield  {author} {\bibinfo {author} {\bibfnamefont {A.~A.}\ \bibnamefont
  {Holmes}}, \bibinfo {author} {\bibfnamefont {N.~M.}\ \bibnamefont {Tubman}},
  \ and\ \bibinfo {author} {\bibfnamefont {C.~J.}\ \bibnamefont {Umrigar}},\
  }\href@noop {} {\bibfield  {journal} {\bibinfo  {journal} {J. Chem. Theory
  Comput.}\ }\textbf {\bibinfo {volume} {12}},\ \bibinfo {pages} {3674}
  (\bibinfo {year} {2016})}\BibitemShut {NoStop}%
\bibitem [{\citenamefont {Sharma}\ \emph {et~al.}(2017)\citenamefont {Sharma},
  \citenamefont {Holmes}, \citenamefont {Jeanmairet}, \citenamefont {Alavi},\
  and\ \citenamefont {Umrigar}}]{S_Sharma2017}%
  \BibitemOpen
  \bibfield  {author} {\bibinfo {author} {\bibfnamefont {S.}~\bibnamefont
  {Sharma}}, \bibinfo {author} {\bibfnamefont {A.~A.}\ \bibnamefont {Holmes}},
  \bibinfo {author} {\bibfnamefont {G.}~\bibnamefont {Jeanmairet}}, \bibinfo
  {author} {\bibfnamefont {A.}~\bibnamefont {Alavi}}, \ and\ \bibinfo {author}
  {\bibfnamefont {C.~J.}\ \bibnamefont {Umrigar}},\ }\href@noop {} {\bibfield
  {journal} {\bibinfo  {journal} {J. Chem. Theory Comput.}\ }\textbf {\bibinfo
  {volume} {13}},\ \bibinfo {pages} {1595} (\bibinfo {year}
  {2017})}\BibitemShut {NoStop}%
\bibitem [{\citenamefont {Chien}\ \emph {et~al.}(2018)\citenamefont {Chien},
  \citenamefont {Holmes}, \citenamefont {Otten}, \citenamefont {Umrigar},
  \citenamefont {Sharma},\ and\ \citenamefont {Zimmerman}}]{S_Chien2018}%
  \BibitemOpen
  \bibfield  {author} {\bibinfo {author} {\bibfnamefont {A.~D.}\ \bibnamefont
  {Chien}}, \bibinfo {author} {\bibfnamefont {A.~A.}\ \bibnamefont {Holmes}},
  \bibinfo {author} {\bibfnamefont {M.}~\bibnamefont {Otten}}, \bibinfo
  {author} {\bibfnamefont {C.~J.}\ \bibnamefont {Umrigar}}, \bibinfo {author}
  {\bibfnamefont {S.}~\bibnamefont {Sharma}}, \ and\ \bibinfo {author}
  {\bibfnamefont {P.~M.}\ \bibnamefont {Zimmerman}},\ }\href@noop {} {\bibfield
   {journal} {\bibinfo  {journal} {J. Phys. Chem. A}\ }\textbf {\bibinfo
  {volume} {122}},\ \bibinfo {pages} {2714} (\bibinfo {year}
  {2018})}\BibitemShut {NoStop}%
\bibitem [{\citenamefont {Hoy}\ and\ \citenamefont {Bunker}(1979)}]{S_Hoy1979}%
  \BibitemOpen
  \bibfield  {author} {\bibinfo {author} {\bibfnamefont {A.~R.}\ \bibnamefont
  {Hoy}}\ and\ \bibinfo {author} {\bibfnamefont {P.~R.}\ \bibnamefont
  {Bunker}},\ }\href@noop {} {\bibfield  {journal} {\bibinfo  {journal} {J.
  Mol. Spectrosc.}\ }\textbf {\bibinfo {volume} {74}},\ \bibinfo {pages} {1}
  (\bibinfo {year} {1979})}\BibitemShut {NoStop}%
\bibitem [{\citenamefont {Schreiber}\ \emph {et~al.}(2008)\citenamefont
  {Schreiber}, \citenamefont {Silva-Junior}, \citenamefont {Sauer},\ and\
  \citenamefont {Thiel}}]{S_Schreiber2008}%
  \BibitemOpen
  \bibfield  {author} {\bibinfo {author} {\bibfnamefont {M.}~\bibnamefont
  {Schreiber}}, \bibinfo {author} {\bibfnamefont {M.~R.}\ \bibnamefont
  {Silva-Junior}}, \bibinfo {author} {\bibfnamefont {S.~P.~A.}\ \bibnamefont
  {Sauer}}, \ and\ \bibinfo {author} {\bibfnamefont {W.}~\bibnamefont
  {Thiel}},\ }\href@noop {} {\bibfield  {journal} {\bibinfo  {journal} {J.
  Chem. Phys.}\ }\textbf {\bibinfo {volume} {128}},\ \bibinfo {pages} {134110}
  (\bibinfo {year} {2008})}\BibitemShut {NoStop}%
\bibitem [{\citenamefont {Daday}\ \emph {et~al.}(2012)\citenamefont {Daday},
  \citenamefont {Smart}, \citenamefont {Booth}, \citenamefont {Alavi},\ and\
  \citenamefont {Filippi}}]{S_Daday2012}%
  \BibitemOpen
  \bibfield  {author} {\bibinfo {author} {\bibfnamefont {C.}~\bibnamefont
  {Daday}}, \bibinfo {author} {\bibfnamefont {S.}~\bibnamefont {Smart}},
  \bibinfo {author} {\bibfnamefont {G.~H.}\ \bibnamefont {Booth}}, \bibinfo
  {author} {\bibfnamefont {A.}~\bibnamefont {Alavi}}, \ and\ \bibinfo {author}
  {\bibfnamefont {C.}~\bibnamefont {Filippi}},\ }\href@noop {} {\bibfield
  {journal} {\bibinfo  {journal} {J. Chem. Theory Comput.}\ }\textbf {\bibinfo
  {volume} {8}},\ \bibinfo {pages} {4441} (\bibinfo {year} {2012})}\BibitemShut
  {NoStop}%
\bibitem [{\citenamefont {Neufeld}\ and\ \citenamefont
  {Thom}(2017)}]{S_Neufeld2017}%
  \BibitemOpen
  \bibfield  {author} {\bibinfo {author} {\bibfnamefont {V.~A.}\ \bibnamefont
  {Neufeld}}\ and\ \bibinfo {author} {\bibfnamefont {A.~J.~W.}\ \bibnamefont
  {Thom}},\ }\href@noop {} {\bibfield  {journal} {\bibinfo  {journal} {J. Chem.
  Phys.}\ }\textbf {\bibinfo {volume} {147}},\ \bibinfo {pages} {194105}
  (\bibinfo {year} {2017})}\BibitemShut {NoStop}%
\end{thebibliography}
\end{document}